\documentclass[aps,prd,amsmath]{revtex4}  %twocolumn,
\usepackage{color,epsfig,graphicx,amssymb,float,multirow,xspace,times}
\definecolor{dblu}{rgb}{0.00,0.00,0.75}
\usepackage[colorlinks,urlcolor=dblu,linkcolor=dblu,citecolor=dblu]{hyperref}
\def \babar {{\it BABAR} }
\begin{document}

      \title{Contributions of the subprocesses $\rho(770,1450,1700)\to K \bar{K}$ and
               $\omega(782,1420,1650)\to K \bar{K}$ for the three-body decays $B\to \eta^{(\prime)} K\bar{K}$}
\author{Ming-Yue Jia$^{1}$}
\author{Jia-Xin Wang$^{1}$}
\author{Li-Fei Yang$^{1}$}
\author{Ai-Jun Ma$^{2}$}     \email{theoma@163.com}
\author{Wen-Fei Wang$^{1}$}\email{wfwang@sxu.edu.cn}

\affiliation{$^1$Institute of Theoretical Physics and \\
                             State Key Laboratory of Quantum Optics and Quantum Optics Devices, \\
                             Shanxi University, Taiyuan, Shanxi 030006, China \\
                   $^2$School of Mathematics and Physics,  Nanjing Institute of Technology,
                             Nanjing, Jiangsu 211167, China
	       }

\date{\today}

%xxxxxxxxxxxxxxxxxxxxxxxxxxxxxxxxxxxxxxxxxxxxxxxxxxxxxxx-abstract-%
\begin{abstract}
As an extension of our prior work, we analyse the resonance contributions for the kaon pair originating from the intermediate $\rho(770)$, $\omega(782)$ and their
excited states in the three-body decays $B\to \eta^{(\prime)} K\bar{K}$ within the perturbative QCD approach.
The information of subprocesses $\rho(770,1450,1700)\to K\bar K$ and $\omega(782,1420,1650)\to K\bar K$ are included in the distribution amplitudes for
 $K\bar K$ system by using the kaon vector time-like form factors.
We calculate the $CP$ averaged branching fractions and the direct $CP$ asymmetries for the relevant quasi-two-body $B$ meson decays.
The branching fractions of the virtual contributions for $K\bar K$ from the Breit-Wigner formula tails of $\rho(770)$ and $\omega(782)$ for these decays are found
comparable to the corresponding contributions from the resonances $\rho(1450,1700)$ and $\omega(1420,1650)$.
Consequently, they constitute a significant component that should be accounted for in the considered three-body decays.
All the predictions in this work are expected to be tested by the LHCb and Belle-II experiments in the future.

\end{abstract}
\maketitle

%%<<<<><><><><><><><><><><><><><><><><><><><><><><><><><><><><><>>>>%%
\section{Introduction}                                                                                                                       %%
\label{sec-intro}                                                                                                                                 %%
%%<<<<><><><><><><><><><><><><><><><><><><><><><><><><><><><><><>>>>%%
%offer us a rich field to investigate
The three-body hadronic $B$ meson decay processes provide us the valuable insights into various phenomena of
the weak and the strong interactions. The $b$-quark weak decay in related processes is described well by the effective
weak Hamiltonian $\mathcal{H}_{\rm eff}$~\cite{rmp68-1125} in Standard Model. While due to the three-body effects
\cite{npps199-341,prd84.094001}, the hadronic interactions and rescattering  processes~\cite{epjc78-897,1512-09284,
prd89.094013,prd71.074016} in their final states, the strong dynamics in these three-body decays is quite complicated
to be captured by a simple expression. The relativistic Breit-Wigner (BW) formula~\cite{BW-model} therefore becomes
a practical way for theorists to describe the scalar, vector, and tensor resonance contributions associated with the
quasi-two-body subprocesses involved in the three-body decays, based on the experimental research with the isobar
formalism~\cite {pr135-B551,pr166.1731, prd11.3165} and Dalitz plot technique~\cite{prd94.1046}.
The quasi-two-body framework based on the perturbative QCD (PQCD) approach~\cite{plb504-6,prd63.054008, prd63.074009,ppnp51-85} has been discussed in detail in~\cite{plb763-29}, which has been followed in Refs.~\cite{2508.09578,prd111.053009,prd111.016002,prd110.036015,epjc84-753,prd109.116009,prd109.056017,
jhep2401-047,cpc46-053104,prd103.056021,prd103.016002,prd102.056017,prd103.013005,prd101.111901,jhep2003-162,
plb791-342,prd95.056008} for the quasi-two-body $B$ meson decays in recent years.

In Ref.~\cite{prd101.111901}, the virtual contribution~\cite{plb25-294,Dalitz62,prd94.072001,plb791-342} for
$K^+K^-$ pair from the BW tail of resonance $\rho(770)^0$, which has been neglected in the experimental
measurement by LHCb Collaboration~\cite{prl123.231802}, was found to be comparable to the contribution of the
subprocess $\rho(1450)^0\to K^+K^-$ in the three-body decays $B^\pm\to \pi^\pm K^+K^-$. Following the work
of~\cite{prd101.111901}, the virtual contributions for kaon pair originating from resonances $\rho(770)$ and
$\omega(782)$ in the three-body $B$ meson decays have been recently studied in Refs.~\cite{prd109.116009,
prd103.056021,prd103.016002,prd110.056001,prd107.116023}. At first sight, the natural modes of $\rho(770)$
and $\omega(782)$ decay into kaon pair will be blocked because of the pole masses for $\rho(770)$ and
$\omega(782)$ which are  apparently below the threshold of the two kaons. But in the processes
$e^+e^- \to K^+K^-$~\cite{pl99b-257,pl107b-297,plb669-217,prd76.072012,zpc39-13,prd88.032013,prd94.112006,
plb779-64,prd99.032001}, $e^+ e^- \to K^0_{S}K^0_{L}$~\cite{pl99b-261,prd63.072002,plb551-27,plb760-314,
prd89.092002,jetp103-720}, $\bar p p \to K^+K^-\pi^0$~\cite{plb468-178,epjc80-453},  $\pi^-p\to K^-K^+n$ and
$\pi^+n\to K^-K^+p$~\cite{prd15.3196,prd22.2595}, the virtual contributions from the BW tail effect of intermediate
state $\rho(770)$ were found to be indispensable for the explanations of the experimental data.
Besides, the resonance $\rho(770)$ is also an important intermediate state for the hadronic $\tau$ decays with kaon
pair~\cite{prd98.032010,prd89.072009,prd53.6037,epjc79-436} and $B$ or $D$ meson decays with $\omega\pi$
pair~\cite{prd92.012013,jhep2401-047,prd107.052010,2502.11159} in the final states.

\begin{figure}[tbp]
\centerline{\epsfxsize=5.5cm \epsffile{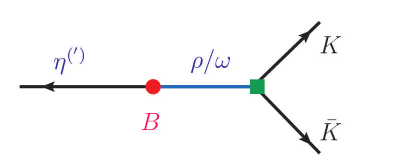}}
\vspace{-0.5cm}
\caption{Schematic view of the cascade decays $B\to\eta^{(\prime)} \rho/\omega \to \eta^{(\prime)}K\bar{K}$,
              where $\rho/\omega$ stands for the intermediate states $\rho(770,1450,1700)$ or $\omega(782,1420,1650)$
              which will decay into kaon pair in this work. }

\label{fig-1}
\vspace{-0.2cm}
\end{figure}

The resonance contributions for the kaon pair originating from the intermediate $\rho(770)$, $\omega(782)$ and their
excited states for the decays $B\to \pi K\bar K$ and $B\to K K\bar K$ have been systematically studied
in Ref.~\cite{prd103.056021} within the PQCD approach. But the contributions of the subprocesses $\rho(770,1450,1700)
\to K \bar{K}$ and $\omega(782,1420,1650)\to K \bar{K}$ are still missing for the three-body decays $B\to \eta^{(\prime)} K\bar{K}$.
In this work, we shall assemble the blocks completely. The schematic view for the cascade decays  $B\to\eta^{(\prime)} \rho/\omega \to \eta^{(\prime)}K\bar{K}$,
where $\rho/\omega$ stands for the intermediate states
$\rho(770,1450,1700)$ or $\omega(782,1420,1650)$ which will decay into kaon pair in this work, is shown in
Fig.~\ref{fig-1}. The intermediate states in relevant cascade decays are generated in the hadronization of the quark-antiquark
pair $q\bar{q}^{(\prime)}$, where ${q}^{(\prime)}$ is a $u$- or $d$-quark. The subprocesses of $\rho\to K\bar{K}$ and
$\omega\to K\bar{K}$ which can't be calculated in PQCD will be introduced into the decay amplitudes via the kaon vector
timelike form factors. The kaon vector timelike form factors are related to its electromagnetic form
factors~\cite{prd72.094003}, which have been extensively studied in Refs.~\cite{epjc79-436,prd67.034012,epjc39-41,
epjc49-697,prd81.094014} on theoretical side and been measured in the reactions $e^+e^- \to K^+K^-$~\cite{zpc39-13,
prd76.072012,prd99.032001} and $e^+e^- \to K^+K^-(\gamma)$~\cite{prd88.032013}. A detailed discussion for the
kaon vector timelike form factors as well as their coefficients are found in Ref.~\cite{prd103.056021}.
The analyses for the relevant three-body $B$ meson decays within the symmetries one is referred to Refs.~\cite{plb564-90,prd72.075013,prd72.094031,prd84.056002,plb727-136,plb726-337,prd89.074043,plb728-579,
prd91.014029}, and for the related papers within QCD factorization can be found in Refs.~\cite{prd72.094003,
prd67.034012,2007-08881,jhep2006-073,plb622-207,plb669-102,prd79.094005,prd88.114014,prd89.074025,prd94.094015,
npb899-247,epjc75-536,prd89.094007,prd87.076007,prd102.053006,prd105.093007,epjc78-845,plb820-136537,
prd99.076010}.

%%~~~~~~~~~~~~~~~~~~~~~~Fig.1~~~~~~~~~~~~~~~~~~~~~~~~%%
%%~~~~~~~~~~~~~~~~~~~~~~Fig.1~~~~~~~~~~~~~~~~~~~~~~~~%%

This paper is organized as follows. In Sec.~\ref{sec-2}, we briefly describe the theoretical framework
for the concerned quasi-two-body $B$ meson decays within the PQCD approach and provide the expression for the
differential branching fraction.
In Sec.~\ref{sec-res}, we present our numerical results for the $CP$ averaged branching fractions
and the corresponding $CP$ asymmetries for the relevant quasi-two-body decay processes as well
as some necessary discussions. A summary and the conclusions of this work are given in Sec.~\ref{sec-sum}. The wave
functions for the initial and final states and the factorization formulas for the related decay amplitudes are
collected in the Appendix~\ref{apx-DAs}.

%%<<<<><><><><><><><><><><><><><><><><><><><><><><><><><><><><><>>>>%%
\section{Framework}                                                                                                                         %%
\label{sec-2}                                                                                                                                      %%
%%<<<<><><><><><><><><><><><><><><><><><><><><><><><><><><><><><>>>>%%
For the $B\to\eta^{(\prime)} \rho/\omega \to \eta^{(\prime)}K\bar{K}$ decays, the weak effective Hamiltonian can be specified as~\cite{rmp68-1125}
\begin{eqnarray}
{\mathcal H}_{\text{eff}} = \frac{G_F}{\sqrt{2}} \Bigg\{ V_{ub} V_{ud(s)}^* \Big[ C_1(\mu) O_1(\mu) + C_2(\mu) O_2(\mu) \Big]
- V_{tb} V_{td(s)}^* \Big[ \sum_{i=3}^{10} C_i(\mu) O_i(\mu) \Big] \Bigg\} + \text{H.c.},
\end{eqnarray}
where $G_F$ is the Fermi coupling constant and $V_{ub,uq,tb,tq}$ $(q=d,s)$ represent the CKM matrix elements.
The $C_i(\mu)$ are the Wilson coefficients at the renormalization scale $\mu$. In this work,
we employ the renormalization group evolution of the Wilson coefficients from higher scale to lower scale as specified in Ref.~\cite{prd63.074009},
where the values of the Wilson coefficients at $m_b$ scale are the following
\begin{eqnarray}
 &&C_1=-0.27034,~~C_2 = 1.11879,~~C_3 = 0.01261,~~C_4 = -0.02695,~~C_5 = 0.00847, \\
 &&C_6=-0.03260,~~C_7 = 0.00109,~~C_8 = 0.00040,~~C_9 = -0.00895,~~C_{10} = 0.00216.
\end{eqnarray}
For the local four-quark operators $O_i$, one has~\cite{rmp68-1125}
\begin{enumerate}
\item[(1)]{Current-current operators}
\begin{eqnarray}
{\renewcommand\arraystretch{1.5}
\begin{array}{ll}
\displaystyle
O_1\, =\,
(\bar{q}_\alpha u_\beta)_{V-A}(\bar{u}_\beta b_\alpha)_{V-A}\;,
& \displaystyle
O_2\, =\, (\bar{q}_\alpha u_\alpha)_{V-A}(\bar{u}_\beta b_\beta)_{V-A}\;;
\end{array}}
\label{eq:operators-1}
\end{eqnarray}
\item[(2)]{ QCD penguin operators}
\begin{eqnarray}
{\renewcommand\arraystretch{1.5}
\begin{array}{ll}
\displaystyle
O_3\, =\, (\bar{q}_\alpha b_\alpha)_{V-A}\sum_{q'}(\bar{q}'_\beta q'_\beta)_{V-A}\;,
& \displaystyle
O_4\, =\, (\bar{q}_\alpha b_\beta)_{V-A}\sum_{q'}(\bar{q}'_\beta q'_\alpha)_{V-A}\;,
\\
\displaystyle
O_5\, =\, (\bar{q}_\alpha b_\alpha)_{V-A}\sum_{q'}(\bar{q}'_\beta q'_\beta)_{V+A}\;,
& \displaystyle
O_6\, =\, (\bar{q}_\alpha b_\beta)_{V-A}\sum_{q'}(\bar{q}'_\beta q'_\alpha)_{V+A}\;;
\end{array}}
\label{eq:operators-2}
\end{eqnarray}
\item[(3)]{ Electroweak penguin operators}
\begin{eqnarray}
{\renewcommand\arraystretch{1.5}
\begin{array}{ll}
\displaystyle
O_7\, =\,
\frac{3}{2}(\bar{q}_\alpha b_\alpha)_{V-A}\sum_{q'}e_{q'}(\bar{q}'_\beta q'_\beta)_{V+A}\;,
& \displaystyle
O_8\, =\,
\frac{3}{2}(\bar{q}_\alpha b_\beta)_{V-A}\sum_{q'}e_{q'}(\bar{q}'_\beta q'_\alpha)_{V+A}\;,
\\
\displaystyle
O_9\, =\,
\frac{3}{2}(\bar{q}_\alpha b_\alpha)_{V-A}\sum_{q'}e_{q'}(\bar{q}'_\beta q'_\beta)_{V-A}\;,
& \displaystyle
O_{10}\, =\,
\frac{3}{2}(\bar{q}_\alpha b_\beta)_{V-A}\sum_{q'}e_{q'}(\bar{q}'_\beta q'_\alpha)_{V-A}\;.
\end{array}}
\label{eq:operators-3}
\end{eqnarray}
\end{enumerate}
The subscripts $\alpha, \ \beta$ are the color indices and the notations
$(\bar{q}'q')_{V\pm A}$ represent $\bar q' \gamma_\mu (1\pm \gamma_5)q'$ with the index $q'=u,\;d,\;s$.

The theoretical analyses of $B$ meson decays are typically performed in the $B$ meson rest frame~\cite{Li:1994zm,Scora:1989ys,Hurth:2025neo,FermilabLattice:2021cdg,Beneke:2021jhp,Boer:2016iez}.
This is because working in this frame provides a clear physical picture in which the final-state particles move fast in different directions,
and eliminates the effects of $B$-meson's intrinsic momentum in the calculations of the decay amplitudes and the integrals of phase space.
Meanwhile, it is always convenient to define the momenta for the initial and final states in the light-cone coordinates for the $B$ meson decays in PQCD approach~\cite{plb504-6,prd63.054008, prd63.074009,ppnp51-85}.
Then, the momentum of $B^+, B^0$ or $B^0_s$ for the concerned decays in this work with the mass $m_B$ can be written as
\begin{eqnarray}
p_B=\frac{m_B}{\sqrt2}(1,1,0_{\rm T})
\end{eqnarray}
in the $B$ meson rest frame. In return, one has the momenta
\begin{eqnarray}
  p_3&=&\frac{m_B}{\sqrt2}(1-\zeta, 0, 0_{\rm T}), \\
  p&=&\frac{m_B}{\sqrt 2}(\zeta, 1, 0_{\rm T})
\end{eqnarray}
for the bachelor state $\eta^{(\prime)}$ and the $K\bar K$ system originating from the resonances $\rho$, $\omega$
or their excited states, respectively. For the spectator quarks in the $B$ meson, $\eta^{(\prime)}$ and the intermediate
states, one has their momenta $k_B=(\frac{m_B}{\sqrt2}x_B, 0, k_{B{\rm T}})$,
$k_3=(\frac{m_B}{\sqrt2}(1-\zeta)x_3, 0, k_{3{\rm T}})$ and $k=(0, \frac{m_B}{\sqrt 2}x, k_{\rm T})$ accordingly.
In addition, one also need to write down a longitudinal polarization vector
\begin{eqnarray}
\epsilon_L=\frac{1}{\sqrt 2}(-\sqrt\zeta, 1/\sqrt\zeta, 0_{\rm T})
\end{eqnarray}
for the resonances involved, which satisfies the relation $ \epsilon_L \cdot p= 0$.
The fractions $x_B$, $x$ and $x_3$ in the momenta above will run from zero to one in
the numerical calculation of this work. It's easy to check that the variable $\zeta=s/m^2_B$ when one defines the
invariant mass square $s=m^2_{K\bar K}\equiv p^2$.

For the subprocesses $\rho(770,1450,1700)\to K \bar{K}$ and $\omega(782,1420,1650)\to K \bar{K}$, the corresponding
electromagnetic form factors with the components of $\rho, \omega$ and $\phi$ resonances and their excited states are written
as~\cite{epjc39-41}
\begin{eqnarray}
  F_{K^+}(s)=&+&\frac12\sum_{\iota=\rho,\rho^\prime,...} c^K_\iota {\rm BW}_\iota(s)
                    +\frac16\sum_{\varsigma=\omega,\omega^\prime,...} c^K_\varsigma {\rm BW}_\varsigma(s)    \nonumber\\
                   &+&\frac13\sum_{\kappa=\phi,\phi^\prime,..} c^K_\kappa {\rm BW}_\kappa(s),  \quad
  \label{def-F-K+}   \\
  F_{K^0}(s)=&-&\frac12\sum_{\iota=\rho,\rho^\prime,...} c^K_\iota {\rm BW}_\iota(s)
                    +\frac16\sum_{\varsigma=\omega,\omega^\prime,...} c^K_\varsigma {\rm BW}_\varsigma(s)    \nonumber\\
                   &+&\frac13\sum_{\kappa=\phi,\phi^\prime,..} c^K_\kappa {\rm BW}_\kappa(s),  \quad
  \label{def-F-K0}
\end{eqnarray}
which are defined by~\cite{epjc39-41}
\begin{eqnarray}
  \langle K^+(p_1) K^-(p_2) | j^{em}_\mu | 0 \rangle &=& (p_1-p_2)_\mu \,F_{K^+}(s), \\
  \langle K^0(p_1)\bar K^0(p_2) | j^{em}_\mu | 0 \rangle &=& (p_1-p_2)_\mu\,F_{K^0}(s),
\end{eqnarray}
where $\rho^\prime, \omega^{\prime}$ stand for $\rho(1450), \omega(1420)$, and the electromagnetic current
$j^{em}_\mu=\frac23\bar u\gamma_\mu u-\frac13\bar d\gamma_\mu d-\frac13\bar s\gamma_\mu s$ carried by the
light quarks $u, d$ and $s$~\cite{npb250-517}. One should note that the state $\phi$ and its excited states are embodied
in components for the form factors $F_{K^+}(s)$ and $F_{K^0}(s)$, but we are not concerned about the subprocess
$\phi(1020)\to K\bar{K}$ in this work, because this decay is a natural decay mode comparing with $\rho(770)$ and
$\omega(782)$ decay into $K\bar{K}$.

The electromagnetic form factors $F_{K^+}$ and $F_{K^0}$ can be separated into the isospin $I=0$ and
$I=1$ components as $F_{K^{+(0)}}=F_{K^{+(0)}}^{I=1} + F_{K^{+(0)}}^{I=0}$, with the
$F_{K^+}^{I=0}=F_{K^0}^{I=0}$ and $F_{K^+}^{I=1}=-F_{K^0}^{I=1}$, and one has
$\langle K^+(p_1) \bar{K}^0(p_2) | \bar u \gamma_\mu d | 0 \rangle =(p_1-p_2)_\mu 2F_{K^+}^{I=1}(s)$~\cite{epjc39-41,
prd96.113003}. Considering only the contribution for ${K^+K^-}$ and ${K^0\bar K^0}$ from the resonant states $\iota=\rho(770,1450,1700)$
and $\varsigma=\omega(782,1420,1650)$, we have~\cite{prd72-094003}
\begin{eqnarray}
  F_{K^+K^-}^{u}(s)&=&F_{K^0\bar K^0}^{d}(s)= +\frac12\sum_{\iota} c^K_\iota {\rm BW}_\iota(s)
                    +\frac12\sum_{\varsigma} c^K_\varsigma {\rm BW}_\varsigma(s),   \label{def-F-u}  \\
  F_{K^+K^-}^{d}(s)&=&F_{K^0\bar K^0}^{u}(s)=-\frac12\sum_{\iota} c^K_\iota {\rm BW}_\iota(s)
                    +\frac12\sum_{\varsigma} c^K_\varsigma {\rm BW}_\varsigma(s).   \label{def-F-d}
\end{eqnarray}
For the $K^+\bar K^0$ and $K^0K^-$ pairs in the final states of the concerned decays in this work which
will have no contribution from the neutral resonances $\omega(782,1420,1650)$,
one has~\cite{prd67.034012,epjc39-41,epjc79-436}
\begin{eqnarray}
  F_{K^+\bar K^0}(s)=F_{K^0K^-}(s)&=&F_{K^+}(s)-F_{K^0}(s) =\sum_{\iota} c^K_\iota {\rm BW}_\iota(s),
   \label{def-F-ud}
\end{eqnarray}
where ${\iota}$ is the only component of $\rho$ family resonances.

The coefficient $c^K_R$ in $F_{K^{+,0}}(s)$ is proportional to the coupling constant $g_{R K\bar K}$, where
$R$ is a resonance $\rho(770,1450,1700)$ or $\omega(782,1420,1650)$ in this work. The values for the related
coefficients have been discussed in detail in Ref.~\cite{prd103.056021}. In the numerical calculation of this work,
we adopt the values
\begin{eqnarray}
    c^K_{\rho(770)}&=&1.247\pm0.019,            \nonumber\\
    c^K_{\omega(782)}&=&1.113\pm0.019,       \nonumber\\
    c^K_{\rho(1450)}&=&-0.156\pm0.015,         \nonumber\\
    c^K_{\omega(1420)}&=&-0.117\pm0.013,    \nonumber\\
    c^K_{\rho(1700)}&=&-0.083\pm0.019,         \nonumber\\
    c^K_{\omega(1650)}&=&-0.083\pm0.019,
    \label{ceffs}
\end{eqnarray}
as they are in~\cite{prd103.056021} for the coefficients in the form factors $F_{K^{+}}(s)$ and $F_{K^{0}}(s)$.

The BW formula in $F_{K^{+,0}}(s)$ has the
form~\cite{zpc48-445,prd101.012006}
\begin{eqnarray}
   {\rm BW}_R= \frac{m_{R}^2}{m_R^2-s-i m_R \Gamma_{R}(s)}\,,
   \label{eq-BW}
\end{eqnarray}
where $R$ means the corresponding intermediate state and the $s$-dependent width is given by
\begin{eqnarray}\label{def-width}
 \Gamma_{R}(s)
             =\Gamma_R\frac{m_R}{\sqrt s} \frac{ \left| \overrightarrow{q} \right|^3}{ \left| \overrightarrow{q_0}\right|^3}
                X^2(\left| \overrightarrow{q} \right| r^R_{\rm BW}).
  \label{eq-sdep-Gamma}
\end{eqnarray}
The Blatt-Weisskopf barrier factor~\cite{BW-X} is given by
\begin{eqnarray}
     X(z)=\sqrt{\frac{1+z^2_0}{1+z^2}},
\end{eqnarray}
with the barrier radius $r^R_{\rm BW}=4.0$ GeV$^{-1}$  as in Refs.~\cite{prd101.012006,prd74-099903,prd92-012012,prd90-072003,prd91-092002}. The magnitude of the momentum
\begin{eqnarray}
   \left| \overrightarrow{q} \right|&=&\frac{1}{2\sqrt s}\sqrt{\left[s-(m_K+m_{\bar K})^2\right]
                               \left[s-(m_K-m_{\bar K})^2\right]}\,,
   \label{def-q}
\end{eqnarray}
and the $\left| \overrightarrow{q_0}\right|$ is $\left| \overrightarrow{q} \right|$ at $s=m^2_R$.

For the $CP$ averaged differential branching fraction ($\mathcal B$) of the concerned quasi-two-body decays, we have the
formula~\cite{prd101.111901,prd79.094005,PDG-2024}
\begin{eqnarray}
 \frac{d{\mathcal B}}{d\zeta}=\tau_B\frac{\left| \overrightarrow{q} \right|^3 \left| \overrightarrow{q_h}\right|^3}
                                                                 {12\pi^3m^5_B}\overline{|{\mathcal A}|^2}\;,
    \label{eqn-diff-bra}
\end{eqnarray}
with $\tau_B$ the mean lifetime of $B$ meson.
The factorization formula for the decay amplitude ${\mathcal A}$ for the quasi-two-body decays
$B\to \eta^{(\prime)} [\rho/\omega\to] K\bar K$ is written as~\cite{plb561-258,prd89.074031}
\begin{eqnarray}
    {\mathcal A}=\phi_B \otimes {\mathcal H} \otimes  \phi^{P\text{-wave}}_{K\bar K} \otimes \phi_{h}
          \label{def-DA-Q2B}
\end{eqnarray}
in the PQCD appraoch. The hard kernel ${\mathcal H}$ contains only one hard gluon exchange at leading order in the
strong coupling $\alpha_s$ according to the Fig.~\ref{fig-feyndiag} in the Appendix~\ref{apx-DAs}.
The symbol $\otimes$ means convolutions in parton
momenta. The distribution amplitudes $\phi_B$, $\phi_{h}$ and $\phi^{P\text{-wave}}_{K\bar K}$ and their their input parameters,  as well as the Lorentz invariant decay amplitudes for the decays
$B\to \eta^{(\prime)} [\rho/\omega\to] K\bar K$   %are given in the Appendix. %~\ref{sec-decayamp}.
are attached in the Appendix~\ref{apx-DAs}.  %~\ref{sec-DAs}.
The magnitude of the momentum ${\small|\overrightarrow{q_{h}}|}$
for the meson $h=\eta^{(\prime)}$ in the rest frame of the resonance is written as
\begin{eqnarray}
   \left| \overrightarrow{q_h}\right|=\frac{1}{2\sqrt s} \sqrt{\left[m^2_{B}-(\sqrt s+m_{h})^2\right]\left[m^2_{B}-(\sqrt s-m_{h})^2\right]},
                \label{def-qh}
\end{eqnarray}
where $m_h$ is the mass for the bachelor meson $\eta^{(\prime)}$. When we face the meson pairs $K^+K^-$ and
$K^0\bar{K}^0$ in the final states, the Eq.~(\ref{def-q}) has a simpler form
\begin{eqnarray}
 \left| \overrightarrow{q} \right|&=&\frac{1}{2}\sqrt{s-4m_K^2}\,.
\end{eqnarray}
The direct $CP$ asymmetry ${\mathcal A}_{CP}$ for the decays in this work is defined as
\begin{eqnarray}
   {\mathcal A}_{CP}=\frac{{\mathcal B}(\bar B\to \bar f)-{\mathcal B}(B\to f)}{{\mathcal B}(\bar B\to \bar f)+{\mathcal B}(B\to f)}.
\end{eqnarray}

%%<<<<><><><><><><><><><><><><><><><><><><><><><><><><><><><><><>>>>%%
\section{Numerical results and Discussions}                                                                                    %%
\label{sec-res}                                                                                                                                    %%
%%<<<<><><><><><><><><><><><><><><><><><><><><><><><><><><><><><>>>>%%

In this work, we adopt the values $f_B=0.190$ GeV and $f_{B_s}=0.230$ GeV for the decay constants of the
$B^{\pm,0}$ and $B^0_s$ mesons~\cite{PDG-2024}, respectively, in the numerical calculation. For these $B$ mesons,
their mean lifetimes are $\tau_{B^\pm}=1.638\times 10^{-12}$~s, $\tau_{B^0}=1.517\times 10^{-12}$~s and
$\tau_{B^0_s}=1.520\times 10^{-12}$~s~\cite{PDG-2024}. The other inputs for the numerical results are presented
in Table~\ref{params}.

%\vspace{-0.2cm}  %%%%%%%%- Table I -%%%
\begin{table*}[tbp]
\centering
\caption{Inputs from~\cite{PDG-2024}: the masses for the relevant particles, the full widths for $\rho(770,1450,1700)$
              and $\omega(782,1420,1650)$ (in units of GeV) and the Wolfenstein parameters.}
\label{params}
\setlength{\tabcolsep}{20pt}
\begin{tabular}{l c c r }\hline\hline
  $m_{B^{\pm}}=5.279$          &  \multicolumn{2}{c}{$m_{B^{0}}=5.280$}          & $m_{B^0_s}\,=5.367$        \\
   $m_{\pi^{\pm}}=0.140$        &  \multicolumn{2}{c}{$m_{\pi^0}\,=0.135$}          &  $m_{K^{\pm}}=0.494$     \\
   $m_{K^{0}}=0.498$             &  \multicolumn{2}{c}{$m_{\eta}=0.548$}               & $m_{\eta^{'}} =0.958$         \\
 \hline
  \multicolumn{2}{l}{$m_{\rho(770)}=0.775$ }
                        &  \multicolumn{2}{r}{$\Gamma_{\rho(770)}=0.147$}                      \\
  \multicolumn{2}{l}{$m_{\omega(782)}=0.783$ }
                       &  \multicolumn{2}{r}{$\Gamma_{\omega(782)}=0.00868$}               \\
  \multicolumn{2}{l}{$m_{\rho(1450)}=1.465\pm0.025$ }
                        &  \multicolumn{2}{r}{$\Gamma_{\rho(1450)}=0.400\pm0.060$}       \\
  \multicolumn{2}{l}{$m_{\omega(1420)}=1.410\pm0.060$ }
                        &  \multicolumn{2}{r}{$\Gamma_{\omega(1420)}=0.290\pm0.190$}    \\
  \multicolumn{2}{l}{$m_{\rho(1700)}=1.720\pm0.020$ }
                        &  \multicolumn{2}{r}{$\Gamma_{\rho(1700)}=0.250\pm0.100$}          \\
  \multicolumn{2}{l}{$m_{\omega(1650)}=1.670\pm0.030$ }
                        &  \multicolumn{2}{r}{$\Gamma_{\omega(1650)}=0.315\pm0.035$}    \\
 \hline
  \multicolumn{2}{l}{$\lambda=0.22501\pm 0.00068$}
                        &  \multicolumn{2}{r}{$A=0.826^{+0.016}_{-0.015}$}                           \\
  \multicolumn{2}{l}{$\bar{\rho} = 0.1591\pm0.0094$}
                        &  \multicolumn{2}{r}{$\bar{\eta}= 0.3523^{+0.0073}_{-0.0071}$}        \\
 \hline\hline
\end{tabular}
\end{table*}
%\vspace{-0.2cm}  %%%%%%%%- Table I -%%%

Utilizing the decay amplitudes collected in the Appendix~\ref{apx-DAs} and the differential branching fractions in the Eq.~(\ref{eqn-diff-bra}), it's trivial to obtain the $CP$ averaged branching fractions and the direct $CP$ asymmetries
for the concerned quasi-two-body decay processes $B\to \eta^{(\prime)}\rho(770,1450,1700)\to \eta^{(\prime)}K\bar K$
and $B\to\eta^{(\prime)}\omega(782,1420,1650)  \to \eta^{(\prime)}K\bar K$ as shown in Tables~\ref{Res-770},
\ref{Res-1450} and \ref{Res-1700}. Because of the small mass difference between $K^\pm$ and $K^0$, the branching
fractions or direct $CP$ asymmetries for the specific quasi-two-body decays
$B\to \eta^{(\prime)}R\to \eta^{(\prime)}K^0\bar K^0$ and $B\to \eta^{(\prime)}R\to \eta^{(\prime)}K^+K^-$ will
be very close,  where the intermediate state $R$ is in  these resonances $\rho(770,1450,1700)^0$ and
$\omega(782,1420,1650)$. Then we omit the corresponding results for those quasi-two-body decays with the subprocesses
$\rho(770, 1450, 1700)^0\to K^0\bar K^0$ and $\omega(782, 1420, 1650)\to K^0\bar K^0$ in Tables~\ref{Res-770}, \ref{Res-1450} and \ref{Res-1700}.  We need to stress here that, the $K^0\bar K^0$ in the final states of the concerned
decays in this work with the $P$-wave resonant origin can not produce the $K^0_SK^0_S$ or $K^0_LK^0_L$ pairs.
%because of the Bose-Einstein statistics.

%%%%%%%%%%%%%-table-rho770-%%
\begin{table}[thb]   %%[H] %%[thb]
\begin{center}
\caption{PQCD predictions of the $CP$ averaged branching fractions and the direct $CP$ asymmetries for the
               quasi-two-body $B\to \eta^{(\prime)}\rho(770) \to \eta^{(\prime)}K\bar K$
               and $B\to \eta^{(\prime)}\omega(782) \to \eta^{(\prime)}K\bar K$ decays.
               The decays with the subprocess $\rho(770)^0\to K^0\bar K^0$ or $\omega(782)\to K^0\bar K^0$
               have the same results as their corresponding decay modes with $\rho(770)^0\to K^+K^-$ or
               $\omega(782)\to K^+K^-$. }
\label{Res-770}   %% ^{+}_{-}
%\small %\footnotesize
 \begin{tabular}{l c r} \hline\hline
 \quad Decay modes       & \qquad${\mathcal B}$\qquad\qquad    &  ${\mathcal A}_{CP}$\quad\quad\;\\
 \hline  %
 $B^+   \to  \eta [\rho(770)^+ \to] K^+ \bar K^0$\;  	
	&$8.03^{+1.69+1.75+0.47+0.25+0.40+0.19}_{-2.31-2.09-0.47-0.24-0.35-0.10}\times10^{-8}$
			&$0.00^{+0.00+0.01+0.01+0.00+0.00+0.00}_{-0.00-0.02-0.01-0.00-0.00-0.00}$    \\
 $B^+   \to  \eta{'} [\rho(770)^+ \to] K^+ \bar K^0$\;
	&$5.01^{+1.07+0.99+0.41+0.15+0.24+0.02}_{-1.46-1.28-0.41-0.15-0.22-0.07}\times10^{-8}$
			&$0.23^{+0.00+0.04+0.01+0.00+0.00+0.32}_{-0.01-0.03+0.01+0.00+0.00+0.32}$  \\
  \hline	
 $B^0   \to  \eta [\rho(770)^0 \to] K^+ K^-$\;
	&$1.51^{+0.26+0.95+0.09+0.05+0.07+0.17}_{-0.34-1.47-0.09-0.05-0.06-0.21}\times10^{-9}$
			&$0.04^{+0.01+0.04+0.02+0.00+0.00+0.06}_{-0.02-0.03-0.02-0.00-0.00-0.06}$    \\
 $B^0   \to  \eta{'} [\rho(770)^0 \to] K^+ K^-$\;
	&$1.03^{+0.18+0.34+0.08+0.04+0.05+0.11}_{-0.24-0.90-0.07-0.03-0.04-0.12}\times10^{-9}$
			&$0.64^{+0.00+0.39+0.02+0.00+0.00+0.43}_{-0.00-0.03-0.02-0.00-0.00-0.47}$    \\
 $B^0   \to  \eta [\omega(782) \to] K^+ K^-$\;
	&$1.41^{+0.22+0.30+0.08+0.05+0.07+0.19}_{-0.28-0.42-0.08-0.05-0.06-0.34}\times10^{-9}$
			&$0.43^{+0.00+0.27+0.01+0.00+0.01+0.59}_{-0.00-0.20-0.01-0.00-0.01-0.60}$    \\
 $B^0   \to  \eta{'}  [\omega(782) \to] K^+ K^- $\;
	&$9.36^{+1.57+3.39+1.03+0.32+0.46+1.14}_{-2.13-1.83-0.38-0.32-0.43-2.09}\times10^{-10}$
			&$0.11^{+0.01+0.15+0.01+0.00+0.00+0.15}_{-0.00-0.37-0.01-0.00-0.00-0.15}$    \\
  \hline
 $B_s^0   \to \eta [\rho(770)^0 \to] K^+ K^-$\;
	&$3.16^{+0.92+0.24+0.33+0.10+0.13+0.28}_{-1.43-0.16-0.31-0.10-0.12-0.47}\times10^{-10}$
			&$-0.24^{+0.01+0.04+0.02+0.00+0.00+0.33}_{-0.02-0.04-0.02-0.00-0.00-0.33}$    \\
 $B_s^0   \to \eta{'} [\rho(770)^0 \to] K^+ K^-$\;
	&$8.76^{+2.15+0.74+0.31+0.27+0.37+0.89}_{-3.17-0.73-0.29-0.27-0.35-1.34}\times10^{-10}$
			&$0.12^{+0.01+0.02+0.00+0.00+0.00+0.16}_{-0.01-0.02-0.00-0.00-0.00-0.16}$    \\
 $B_s^0   \to \eta [\omega(782) \to] K^+ K^-$\;
	&$3.08^{+0.90+0.22+0.32+0.11+0.13+0.28}_{-1.37-0.17-0.30-0.10-0.12-0.46}\times10^{-10}$
			&$-0.25^{+0.01+0.04+0.02+0.00+0.00+0.34}_{-0.02-0.04-0.02-0.00-0.00-0.34}$    \\
 $B_s^0   \to \eta{'}  [\omega(782) \to] K^+ K^- $\;
	&$8.46^{+2.08+0.71+0.29+0.29+0.36+0.86}_{-3.07-0.71-0.29-0.28-0.33-1.29}\times10^{-10}$
			&$0.12^{+0.01+0.02+0.00+0.00+0.00+0.16}_{-0.01-0.02-0.00-0.00-0.00-0.16}$    \\
\hline\hline
\end{tabular}
\end{center}
\end{table}
%%%%%%%%%%%%%-table-rho770-%%

%%%%%%%%%%%%%-table-rho1450-%%
\begin{table}[thb]   %%[H] %%[thb]
\begin{center}
\caption{PQCD predictions of the $CP$ averaged branching fractions and the direct $CP$ asymmetries for the
               quasi-two-body $B\to \eta^{(\prime)}\rho(1450) \to \eta^{(\prime)}K\bar K$
               and $B\to \eta^{(\prime)} \omega(1420) \to \eta^{(\prime)}K\bar K$ decays.
               Those decays with the subprocess $\rho(1450)^0\to K^0\bar K^0$ or $\omega(1420)\to K^0\bar K^0$
               have the same results as their corresponding decay modes with $\rho(1450)^0\to K^+K^-$ or
               $\omega(1420)\to K^+K^-$. }
\label{Res-1450}   %% ^{+}_{-}
%\small %\footnotesize
 \begin{tabular}{l c r} \hline\hline
 \quad Decay modes       & \qquad${\mathcal B}$\qquad\qquad    &  ${\mathcal A}_{CP}$\quad\quad\;\\
 \hline  %
 $B^+   \to  \eta [\rho(1450)^+ \to] K^+ \bar K^0$\;  \qquad 	
	&\qquad$5.49^{+1.18+1.13+0.32+1.00+0.26+0.27}_{-1.62-1.34-0.33-1.11-0.24-0.16}\times10^{-8} \qquad$
		&$0.01^{+0.00+0.01+0.01+0.00+0.00+0.01}_{-0.00-0.01-0.01-0.00-0.00-0.01}$    \\
 $B^+   \to  \eta{'} [\rho(1450)^+ \to] K^+ \bar K^0$\;
	&$3.47^{+0.77+0.66+0.29+0.02+0.16+0.16}_{-1.03-0.85-0.28-0.03-0.15-0.10}\times10^{-8}$
		&$0.23^{+0.01+0.05+0.01+0.00+0.00+0.02}_{-0.01-0.03-0.01-0.00-0.00-0.10}$   \\
 \hline	
 $B^0   \to  \eta [\rho(1450)^0 \to] K^+ K^-$\;
	&$1.81^{+0.32+1.01+0.10+0.37+0.08+0.22}_{-0.40-1.95-0.11-0.37-0.07-0.19}\times10^{-9}$
		&$0.00^{+0.01+0.02+0.02+0.00+0.00+0.00}_{-0.00-0.02-0.02-0.00-0.00-0.00}$    \\
 $B^0   \to  \eta{'} [\rho(1450)^0 \to] K^+ K^-$\;
	&$1.35^{+0.25+0.46+0.10+0.24+0.06+0.13}_{-0.30-1.49-0.09-0.28-0.05-0.12}\times10^{-9}$
		&$0.65^{+0.00+0.45+0.02+0.00+0.00+0.04}_{-0.00-0.03-0.02-0.00-0.00-0.06}$    \\
 $B^0   \to  \eta [\omega(1420) \to] K^+ K^-$\;
	&$6.00^{+0.83+1.24+0.33+1.21+0.27+0.83}_{-1.05-1.74-0.34-1.37-0.26-2.03}\times10^{-10}$
		&$0.45^{+0.01+0.27+0.01+0.00+0.01+0.11}_{-0.00-0.20-0.01-0.00-0.01-0.09}$    \\
 $B^0   \to  \eta{'}  [\omega(1420) \to] K^+ K^- $\;
	&$4.00^{+0.64+1.15+0.30+0.84+0.20+0.57}_{-0.84-0.88-0.29-0.94-0.19-1.38}\times10^{-10}$
		&$0.13^{+0.01+0.28+0.01+0.00+0.00+0.05}_{-0.00-0.36-0.01-0.00-0.01-0.02}$\\
 \hline
 $B_s^0   \to \eta [\rho(1450)^0 \to] K^+ K^-$\;
	&$4.54^{+1.34+0.39+0.49+0.83+0.54+0.39}_{-2.09-0.30-0.45-0.92-0.60-0.68}\times10^{-10}$
		&$-0.24^{+0.02+0.05+0.03+0.00+0.00+0.05}_{-0.02-0.05-0.03-0.00-0.00-0.06}$    \\
 $B_s^0   \to \eta{'} [\rho(1450)^0 \to] K^+ K^-$\;
	&$1.22^{+0.30+0.09+0.04+0.22+0.05+0.12}_{-0.45-0.09-0.04-0.25-0.05-0.17}\times10^{-9}$
		&$0.14^{+0.01+0.02+0.00+0.00+0.00+0.02}_{-0.01-0.02-0.00-0.00-0.00-0.02}$    \\
 $B_s^0   \to \eta [\omega(1420) \to] K^+ K^-$\;
	&$1.59^{+0.47+0.12+0.16+0.33+0.07+0.14}_{-0.72-0.11-0.16-0.37-0.06-0.24}\times10^{-10}$
		&$-0.24^{+0.01+0.05+0.03+0.00+0.00+0.05}_{-0.02-0.04-0.03-0.00-0.00-0.06}$    \\
 $B_s^0   \to \eta{'}  [\omega(1420) \to] K^+ K^- $\;
	&$4.04^{+1.02+0.29+0.15+0.85+0.17+0.39}_{-1.49-0.28-0.14-0.95-0.16-0.58}\times10^{-10}$
		&$0.15^{+0.01+0.02+0.00+0.00+0.00+0.02}_{-0.01-0.02-0.00-0.00-0.00-0.02}$    \\
\hline\hline
\end{tabular}
\end{center}
\end{table}
%%%%%%%%%%%%%-table-rho1450-%%

%%%%%%%%%%%%%-table-rho1700-%%
\begin{table}[thb]   %%[H] %%[thb]
\begin{center}
\caption{PQCD predictions of the $CP$ averaged branching fractions and the direct $CP$ asymmetries for the
               quasi-two-body $B\to \eta^{(\prime)}\rho(1700) \to \eta^{(\prime)}K\bar K$
               and $B\to \eta^{(\prime)}\omega(1650) \to \eta^{(\prime)}K\bar K$ decays.
               Those decays with the subprocess $\rho(1700)^0\to K^0\bar K^0$ or $\omega(1650)\to K^0\bar K^0$
               have the same results as their corresponding decay modes with $\rho(1700)^0\to K^+K^-$ or
               $\omega(1650)\to K^+K^-$. }
\label{Res-1700}   %% ^{+}_{-}
%\small %\footnotesize
 \begin{tabular}{l c r} \hline\hline
 \quad Decay modes       & \qquad${\mathcal B}$\qquad\qquad    &  ${\mathcal A}_{CP}$\quad\quad\;\\
 \hline  %
 $B^+   \to  \eta [\rho(1700)^+ \to] K^+ \bar K^0$\; \qquad  	
	&\qquad $6.00^{+1.28+1.37+0.36+2.43+0.28+0.38}_{-1.72-1.62-0.35-3.06-0.27-0.25}\times10^{-8} \qquad$
		&$0.00^{+0.00+0.01+0.01+0.00+0.00+0.00}_{-0.00-0.02-0.01-0.00-0.00-0.00}$    \\
 $B^+   \to  \eta{'} [\rho(1700)^+ \to] K^+ \bar K^0$\;
	&$3.78^{+0.81+0.77+0.32+1.53+0.18+0.21}_{-1.11-1.01-0.31-1.93-0.17-0.14}\times10^{-8}$
		&$0.24^{+0.01+0.05+0.01+0.00+0.00+0.02}_{-0.01-0.03-0.01-0.00-0.00-0.02}$   \\
 \hline	
 $B^0   \to  \eta [\rho(1700)^0 \to] K^+ K^-$\;
	&$2.11^{+0.38+1.40+0.12+0.85+0.09+0.30}_{-0.45-2.21-0.13-1.08-0.09-0.27}\times10^{-9}$
		&$-0.02^{+0.01+0.07+0.02+0.00+0.00+0.01}_{-0.01-0.05-0.02-0.00-0.00-0.01}$    \\
 $B^0   \to  \eta{'} [\rho(1700)^0 \to] K^+ K^-$\;
	&$7.89^{+1.44+2.55+0.57+3.20+0.34+0.83}_{-1.81-7.12-0.56-4.01-0.32-0.81}\times10^{-10}$
		&$0.65^{+0.00+0.44+0.02+0.00+0.00+0.03}_{-0.00-0.03-0.02-0.00-0.00-0.06}$  \\
 $B^0   \to  \eta [\omega(1650) \to] K^+ K^-$\;
	&$6.81^{+1.04+1.44+0.39+2.76+0.32+1.12}_{-1.29-2.12-0.38-3.49-0.30-2.66}\times10^{-10}$
		&$0.55^{+0.01+0.29+0.01+0.00+0.01+0.07}_{-0.00-0.20-0.01-0.00-0.01-0.07}$    \\
 $B^0   \to  \eta{'}  [\omega(1650) \to] K^+ K^- $\;
	&$4.82^{+0.88+1.68+0.35+1.96+0.24+0.83}_{-0.98-0.96-0.35-2.45-0.23-1.96}\times10^{-10}$
		&$0.19^{+0.00+0.21+0.01+0.00+0.01+0.06}_{-0.00-0.38-0.01-0.00-0.01-0.04}$    \\
 \hline
 $B_s^0   \to \eta [\rho(1700)^0 \to] K^+ K^-$\;
	&$4.40^{+1.33+0.50+0.52+1.79+0.19+0.37}_{-2.10-0.40-0.48-2.25-0.17-0.57}\times10^{-10}$
		&$-0.27^{+0.03+0.05+0.03+0.00+0.00+0.07}_{-0.03-0.05-0.03-0.00-0.00-0.07}$    \\
 $B_s^0   \to \eta{'} [\rho(1700)^0 \to] K^+ K^-$\;
	&$1.31^{+0.33+0.05+0.04+0.53+0.06+0.12}_{-0.48-0.05-0.04-0.67-0.05-0.18}\times10^{-9}$
		&$0.12^{+0.01+0.02+0.00+0.00+0.00+0.02}_{-0.01-0.02-0.00-0.00-0.00-0.02}$    \\
 $B_s^0   \to \eta [\omega(1650) \to] K^+ K^-$\;
	&$1.54^{+0.47+0.16+0.17+0.78+0.06+0.13}_{-0.73-0.13-0.16-0.62-0.06-0.20}\times10^{-10}$
		&$-0.28^{+0.02+0.05+0.02+0.00+0.00+0.07}_{-0.02-0.05-0.04-0.00-0.00-0.08}$    \\
 $B_s^0   \to \eta{'}  [\omega(1650) \to] K^+ K^- $\;
	&$4.41^{+1.10+0.37+0.15+1.79+0.19+0.41}_{-1.62-0.35-0.15-2.25-0.17-0.61}\times10^{-10}$
		&$0.12^{+0.01+0.02+0.00+0.00+0.00+0.02}_{-0.01-0.02-0.00-0.00-0.00-0.02}$    \\
\hline\hline
\end{tabular}
\end{center}
\end{table}
%%%%%%%%%%%%%-table-rho1700-%%

For these predicted results, their first error is induced from the uncertainties of the shape parameter
$\omega_B = 0.40\pm0.04 $ and $\omega_{B_s} = 0.50\pm 0.05$ for the $B^{+,0} $ and $B^0_s$ meson, the second error comes from the
uncertainties of the Gegenbauer moments $a^0_R=0.25 \pm0.10$, $a^t_R=-0.50\pm0.20$, and $a^s_R=0.75\pm0.25$ for the distribution
amplitudes of the intermediate states in Eqs.~(\ref{def-DA-0})-(\ref{def-DA-s}), respectively.
The $\eta-\eta{'}$ mixing angle $\phi=(40.0\pm2.0_{\rm stat}\pm0.6_{\rm syst})^{\circ}$ contributes the third error.
The fourth and fifth errors originate from the coefficient $c^K_R$ of $F_{K^{+,0}}(s)$ in Eq.~(\ref{ceffs}) and
the Wolfenstein parameters in Table~\ref{params}, respectively.
The last error comes from the hard scale $t$ with the range of $0.75t$ to $1.25t$
 and the QCD scale $\Lambda_{QCD}=0.25\pm0.05$ GeV with reference to ~\cite{prd76.074018}.
There are other errors for the results in this work, which come from the uncertainties of the masses and the decay constants
of the initial and final states, and the chiral scale parameters of the bachelor mesons, etc., are small and have been neglected.

The branching fraction and direct $CP$ asymmetry for the two-body decay $B^+ \to  \eta \rho(770)^+$
were averaged in~\cite{PDG-2024} with the data ${\mathcal{B}}=(7.0\pm2.9)\times 10^{-6}$ and
${\mathcal{A}}_{CP}=0.11\pm0.11$ from the original results
${\mathcal{B}}=(9.9\pm1.2\pm0.8)\times 10^{-6}$ and ${\mathcal{A}}_{CP}=0.13\pm0.11\pm0.02$,
${\mathcal{B}}=(4.1^{+1.4}_{-1.3}\pm0.4)\times 10^{-6}$ and ${\mathcal{A}}_{CP}=-0.04^{+0.34}_{-0.32}\pm0.01$
presented by \babar and Belle collaborations in~\cite{prd78.011107} and \cite{prd75.092005}, respectively.
To illustrate the capabilities of the method adopt in this work, we calculate the branching fraction and direct $CP$ asymmetry
for the quasi-two-body decay $B^+ \to  \eta \rho(770)^+ \to \eta \pi^+\pi^0$ with the results
\begin{eqnarray}
    {\mathcal{B}}&=&(6.34^{+1.73}_{-2.20})\times10^{-6},
        \label{br4pipi} \\
    {\mathcal{A}}_{CP}&=&-0.006^{+0.006}_{-0.010},
\end{eqnarray}
where the individual errors have been added in quadrature.
Apparently our predictions are consistent with the corresponding data in~\cite{PDG-2024} for the two-body decay
$B^+ \to  \eta \rho(770)^+$ in view of ${\mathcal{B}}( \rho(770)^+ \to \pi^+\pi^0)\approx100\%$~\cite{PDG-2024}.

The coupling constants $g_{\rho(1450)^0\pi^+\pi^-}$ and $g_{\rho(1450)^0K^+K^-}$ can be achieved by the
the expression %~\cite{2007-02558}
\begin{eqnarray}
  g_{\rho(1450)^0h^+h^-}=\sqrt{\frac{6\pi m^2_{\rho(1450)}\Gamma_{\rho(1450)}{\mathcal B}_{\rho(1450)^0\to h^+h^-}}{q^3}}\,,
  \label{def-cocont}
\end{eqnarray}
with $h$ a pion or kaon, and $q=\frac{1}{2}\sqrt{m^2_{\rho(1450)}-4m_h^2}$. With the help of the relation
$g_{\rho(1450)^0K^+K^-}\approx \frac12 g_{\rho(1450)^0\pi^+\pi^-}$~\cite{epjc39-41}
one has the ratio~\cite{prd101.111901}
\begin{eqnarray}
  R_{\rho(1450)} &=& \frac{{\mathcal B}(\rho(1450)^0\to K^+ K^-)}{{\mathcal B}(\rho(1450)^0 \to\pi^+\pi^-)} %\nonumber\\
                           \approx \frac{g^2_{\rho(1450)^0K^+K^-} (m^2_{\rho(1450)}-4m^2_K)^{3/2}}
                                                {g^2_{\rho(1450)^0\pi^+\pi^-}(m^2_{\rho(1450)}-4m^2_\pi)^{3/2}} \nonumber\\
                         &=&  0.107.
   \label{th-Rho-1450}
\end{eqnarray}   %%% aaaabcd
Since the pole mass of $\rho(770)$ is smaller than that of the Kaon pair, it is not proper to calculate the ratio $R_{\rho(770)}$ through a similar formula.
Considering the smaller phase space available in $\rho(770)\to K\bar K$ relative to $\rho(770)\to \pi\pi$,
along with the ratio $R_{\rho(1450)}$, it's easy to understand the branching fraction in Table~\ref{Res-770}
for the quasi-two-body decay process $B^+\to\eta [\rho(770)^+ \to] K^+ \bar K^0$,
which is two orders smaller than the data in~\cite{PDG-2024} for the two-body decay $B^+ \to\eta\rho(770)^+$.

From the branching fractions in Tables~\ref{Res-770} and \ref{Res-1450}, one can find that the virtual contributions
for the kaon pair from the BW tails of the resonances $\rho(770)$ and $\omega(782)$ are on the order of or larger than the
corresponding contributions from $\rho(1450)$ and $\omega(1420)$. But due to the relatively smaller value for the
$c^K_{\omega(1420)}$ than that for $c^K_{\rho(1450)}$ adopted in this work, the ratio of the branching fractions for
a concerned quasi-two-body decay with the subprocesses $\omega(1420)\to K\bar{K}$ and $\omega(782)\to K\bar{K}$
is obviously smaller than the ratio of that with the $\rho(1450)\to K\bar{K}$ and $\rho(770)\to K\bar{K}$.
While the branching fractions in Table~\ref{Res-1700} with the subprocesses $\rho(1700)\to K\bar{K}$ and
$\omega(1650)\to K\bar{K}$ are comparable to their corresponding branching fractions in Table~\ref{Res-1450}
which come from the contribution of the resonances $\rho(1450)$ and $\omega(1420)$ for the kaon pair.

%%%%%%%%%%%%%%%%%%%%%%=Fig.2
\begin{figure}[tbp]
\centerline{\epsfxsize=13 cm \epsffile{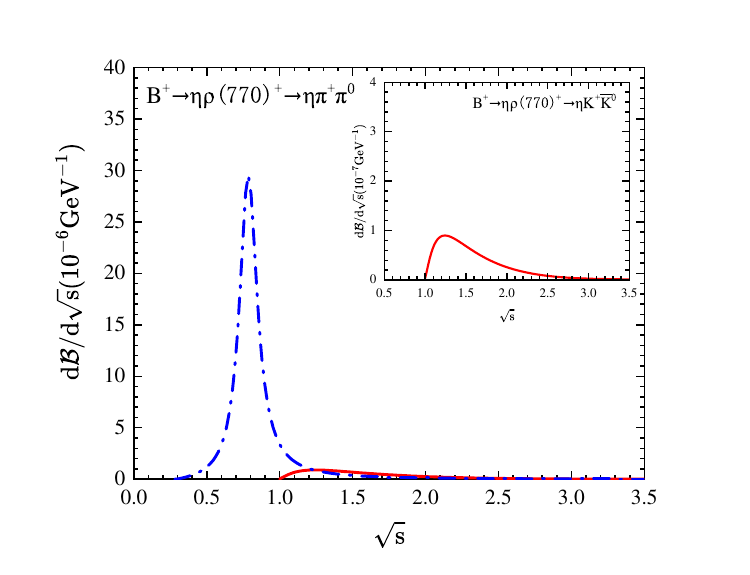}}
\caption{The differential branching fractions for the decays $B^+\to \eta [\rho(770)^+\to] K^+\bar{K}^0$ and
               $B^+\to \eta [\rho(770)^+\to] \pi^+\pi^0$. The big diagram is for the comparison
               for the differential branching fractions of $B^+\to \eta [\rho(770)^+\to] K^+\bar{K}^0$ and
               $B^+\to \eta [\rho(770)^+\to] \pi^+\pi^0$, in which the solid line for $B^+\to \eta [\rho(770)^+\to] K^+\bar{K}^0$
               is magnified by a factor of $10$.
              }
\label{fig-dB-770}
\end{figure}
%%%%%%%%%%%%%%%%%%%%%%=Fig.2

In Fig.~\ref{fig-dB-770}, we show the differential branching fractions for the decays
$B^+\to \eta [\rho(770)^+\to] K^+\bar{K}^0$ and $B^+\to \eta [\rho(770)^+\to] \pi^+\pi^0$. In order to provide good contrast, we magnify the value of each point of the solid line for $B^+\to \eta [\rho(770)^+\to] K^+\bar{K}^0$ by a factor
of $10$ in the big diagram. From the diagrams in Fig.~\ref{fig-dB-770},  one can find that the differential branching fraction
for $B^+\to \eta [\rho(770)^+\to] K^+\bar{K}^0$, which is related to the virtual contribution, does not look like a typical
BW curve for the normal quasi-two-body decay process like $B^+\to \eta [\rho(770)^+\to] \pi^+\pi^0$.
A broad bump is found for the process $B^+\to \eta [\rho(770)^+\to] K^+\bar{K}^0$ with its peak at about
$\sqrt{s}=1.35$ GeV. This bump does not mean the mass for a new resonance. It is actually generated from the BW tail of
$\rho(770)^+$ along with a strong suppression factor $\left| \overrightarrow{q_h}\right|^3$ in Eq.~(\ref{eqn-diff-bra}).
Furthermore, we find that the curve for the differential branching fraction with the virtual contribution involved will not be
effected much by the decay width of the related intermediate state. This may be somewhat surprising.
Acctually the imaginary part of the denominator in the BW formula for $\rho(770)$ (or $\omega(782)$),
which hold the energy dependent width of the intermediate state, will become unimportant when the invariant mass square
for kaon pair is large enough. The BW expression is then charged only by the coefficient $c^K_R$ in the time-like form
factors for kaons in this work as well as the gap between the $s$ and the squared mass $m^2_R$ for the intermediate state.
Although the threshold for kaon pair is not far from the pole masse of $\rho(770)$ (or $\omega(782)$),  due to the strong
suppression from the factor $\left| \overrightarrow{q}\right|^3$ in Eq.~(\ref{eqn-diff-bra}), the differential branching fraction
with the subprocess $\rho(770)^+\to K^+\bar{K}^0$ will reach its peak at about $1.35$ GeV as it is in Fig.~\ref{fig-dB-770}.

In principle, the perturbative QCD calculations are applicable when the energy release is large enough.
It is therefore worthwhile to test the reliability of the PQCD framework in the high-mass range here.
Fortunately, the evolution of the time-like form factor $F_{K\bar K}(s)$ in the decay amplitude ${\cal A}$ and
the phase space factor $| \overrightarrow{q_h}|$ in the differential branching fraction of Eq.~(\ref{eqn-diff-bra})
 will naturally suppress the
resonant contribution from the region where the invariant mass of the kaon pair is much higher than the pole mass of the resonance.
Take $B^+ \to \eta [\rho(770)^+ \to] K^+ \bar K^0$ as an example, the main portion of its branching ratio can be easily found to
lie in the mass region around $1.35$ GeV as shown in Fig.~\ref{fig-dB-770}.
Numerically, the central values of its branching ratio are calculated as $5.69 \times 10^{-8}$ and $6.78 \times 10^{-8}$ when the $K^+ \bar K^0$ invariant mass upper limit is set to $2.0$ GeV and $2.5$ GeV, respectively, which amount
to be $70.86\%$ and $84.43\%$ of the value $8.03 \times 10^{-8}$ as given in Table~\ref{Res-770}.
The similar ratios become $84.88\%$ and $93.81\%$ for $B^+ \to \eta [\rho(1450)^+ \to] K^+ \bar K^0$, and $84.17\%$ and $95.00\%$ for
$B^+ \to \eta [\rho(1700)^+ \to] K^+ \bar K^0$. The results show that for these considered resonance contributions, they primarily originate from the low invariant mass region.
Meanwhile, compared to the contribution for a full resonance, the high invariant mass region has a more significant impact on the contribution for the off-shell state.
On the whole, it indicates the PQCD predictions for the present processes in this work are reliable.

%%<<<<><><><><><><><><><><><><><><><><><><><><><><><><><><><><><>>>>%%
\section{Summary}                                                                                                                            %%
\label{sec-sum}                                                                                                                                  %%
%%<<<<><><><><><><><><><><><><><><><><><><><><><><><><><><><><><>>>>%%

In this work, we studied the contributions of the subprocesses $\rho(770,1450,1700) \to K \bar{K}$ and $\omega(782,1420,1650)\to K \bar{K}$
in the three-body decays $B\to \eta^{(\prime)} K\bar{K}$ by employing the perturbative QCD approach.
The kaon vector timelike form factors for the subprocesses of $\rho\to K\bar{K}$ and $\omega\to K\bar{K}$, which are related to its electromagnetic form factors,
were introduced into the distribution amplitudes for $K\bar K$ system.
The $CP$ averaged branching fractions for the considered decays were predicted to be on the order of $10^{-10}$ to $10^{-8}$.
The contributions for the kaon pair from the Breit-Wigner tails of the resonances $\rho(770)$ and $\omega(782)$ have been found on the order of or larger than the
corresponding contributions from $\rho(1450,1700)$ and $\omega(1420,1650)$, which means that those virtual contributions can not be ignored in the  $K\bar{K}$ system.
We also found that the curve for the differential branching fraction with the virtual contribution involved will not be
effected much by the decay width of the related intermediate state $\rho(770)$ (or $\omega(782)$), since the imaginary part of the denominator in the BW formula
will become unimportant when the invariant mass square for kaon pair is much larger that the pole mass of the resonance.
Since the well-determined relative phase angles between the states of the $\rho$ or $\omega$ family in the $K\bar K$ form factor are currently lacking,
together with the lack of the experimental measurement for the decays considered in this work, we leave the total branching fractions and $CP$ asymmetries
 accounting for all resonant contributions and their interference effects in the concerned decays to future studies.
All the PQCD predictions in this work were expected to be measured
in the future high-statistics experiments by Belle II and LHCb.

%%%~~~~~~~~~~~~~~~~~~~~~~~~~~~~~~~~~~~~~~~Acknowledgments~~~%%%
\begin{acknowledgments}
This work was supported in part by the National Natural Science Foundation of China under Grants
No. 12575100 and No. 12205148, and the Qing Lan Project of Jiangsu Province.
L.F. Yang was also supported in part by the Postgraduate Education
Innovation Program of Shanxi Province under Grant No. 2024KY104 .
\end{acknowledgments}
%%%~~~~~~~~~~~~~~~~~~~~~~~~~~~~~~~~~~~~~~~Acknowledgments~~~%%%

%%<<<<><><><><><><><><><><><><><><><><><><><><><><><><><><><><><>>>>%%
\appendix                                                                                                                                            %%
%%<<<<><><><><><><><><><><><><><><><><><><><><><><><><><><><><><>>>>%%
%%<<<<><><><><><><><><><><><><><><><><><><>%
\section{Distribution amplitudes} \label{apx-DAs}
%%<<<<><><><><><><><><><><><><><><><><><><>%

In three-body decays $B\to \eta^{(\prime)} K\bar{K}$, the mixing between $\eta$ and $\eta^{\prime}$ are
taking into account.  The physical states $\eta$ and $\eta^\prime$ are made from $\eta_q=(u\bar u+d\bar d)/\sqrt2$
and $\eta_s=s\bar s$ at quark level in early studies~\cite{plb449-339,prd58.114006} with
\begin{equation}
\left(\begin{array}{c} \eta \\ \eta^{\prime} \end{array} \right)
                      = \left(\begin{array}{cc}  \cos{\phi} & -\sin{\phi} \\ \sin{\phi} & \cos{\phi} \\ \end{array} \right)
 \left(\begin{array}{c} \eta_q \\ \eta_s \end{array} \right),
 \label{mixing-eta}
\end{equation}
where the mixing angle $\phi=39.3^{\circ}\pm1.0^{\circ}$, and the decay constants $f_{\eta_q}=(1.07\pm0.02)f_\pi$
and $f_{\eta_s}=(1.34\pm0.06)f_\pi$~\cite{plb449-339,prd58.114006}. In this work,
the recent measurement of the angle $\phi=(40.0\pm2.0_{\rm stat}\pm0.6_{\rm syst})^{\circ}$ by BESIII
Collaboration~\cite{prd108.092003} is adopt for the numerical calculation.

The light cone wave functions for the states $h=\eta_q$ and $h=\eta_s$ are written
as~\cite{prd71.014015,jhep9901-010}
\begin{eqnarray}
	\Phi_{h}(p_3,x_3)\equiv \frac{i}{\sqrt{2N_C}}\gamma_5
	\left [{ p \hspace{-2.0truemm}/ }_3 \phi_{h}^{A}(x_3)+m_{03} \phi_{h}^{P}(x_3)
	+ m_{03} ({ n \hspace{-2.2truemm}/ } { v \hspace{-2.2truemm}/ } - 1)\phi_{h}^{T}(x_3)\right ],
\end{eqnarray}
where $m_{03}$ is the corresponding meson chiral mass.
The distribution amplitudes $\phi_{\eta_{q(s)}}^{A,P,T}$ for $\eta_{q(s)}$ are given as~\cite{prd71.014015,jhep9901-010}:
\begin{eqnarray}
	\phi_{\eta_{q(s)}}^A(x) &=&  \frac{f_{q(s)}}{2\sqrt{2N_c} }    6x (1-x)
	                    \bigg[1+a^{\eta}_1C^{3/2}_1(2x-1)+a^{\eta}_2 C^{3/2}_2(2x-1)+a^{\eta}_4C^{3/2}_4(2x-1)\bigg],
	          \nonumber\\
	\phi_{\eta_{q(s)}}^P(x) &=&   \frac{f_{q(s)}}{2\sqrt{2N_c} }
	                             \bigg[ 1+(30\eta_3-\frac{5}{2}\rho^2_{\eta_{q(s)} } )C^{1/2}_2(2x-1)-3\big[\eta_3\omega_3
	                            +\frac{9}{20}\rho^2_{\eta_{q(s)} }(1+6a^{\eta }_2)\big]     \nonumber\\
	                   &\times& C^{1/2}_4(2x-1)\bigg],
	          \nonumber\\
	\phi_{\eta_{q(s)}}^T(x) &=&  \frac{f_{q(s)}}{2\sqrt{2N_c} } (1-2x)
	                \bigg[ 1+6 (5\eta_3-\frac{1}{2}\eta_3\omega_3
	               -\frac{7}{20}\rho^2_{\eta_{q(s)}}-\frac{3}{5}\rho^2_{\eta_{q(s)} }a_2^{\eta} ) (1-10x+10x^2 )\bigg],
\end{eqnarray}
with the Gegenbauer moments~\cite{prd71.014015}
\begin{eqnarray}
	a^{\eta_{q(s)}}_1=0,   \quad a^{\eta_{q(s)}}_2=0.115, \quad a^{\eta_{q(s)}}_4=-0.015,
	 \quad \eta_3=0.015,  \quad \omega_3=-3.0,
\end{eqnarray}
and the paprameters $\rho_{\eta_q}=2m_q/m_{qq}$ for $\eta_q$ and $\rho_{\eta_s}=2m_s/m_{ss}$.
The chiral masses $m_{03}=m_0^q$ for $\eta_q$ and $m_{03}=m_0^s$ for $\eta_s$ are defined as~\cite{prd74.074024}
\begin{eqnarray}
    m_0^q&\equiv& \frac{m_{qq}^2}{2m_q}=\frac{1}{2m_q} \left[m_{\eta}^2\cos^2\phi+m_{\eta'}^2\sin^2\phi-
                \frac{\sqrt{2}f_s}{f_q}(m_{\eta'}^2-m_\eta^2)\cos\phi\sin\phi\right],    \\
     m_0^s&\equiv& \frac{m_{ss}^2}{2m_s}=\frac{1}{2m_s} \left[m_{\eta'}^2\cos^2\phi+m_{\eta}^2\sin^2\phi-
               \frac{f_q}{\sqrt{2}f_s}(m_{\eta'}^2-m_\eta^2)\cos\phi\sin\phi\right].
\end{eqnarray}
The Gegenbauer polynomials $C^{\nu}_n(t)$ ($n=1,2,4$ and $\nu=1/2, 3/2$) above could be found in Ref.~\cite{prd76.074018}.

The $B$ meson light-cone matrix element in Eq.~(\ref{def-DA-Q2B}) is decomposed as~\cite{npb592-3,prd76.074018}
\begin{eqnarray}
\Phi_B=\frac{i}{\sqrt{2N_c}} (p{\hspace{-1.8truemm}/}_B+m_B)\gamma_5\phi_B (k_B),
\label{bmeson}
\end{eqnarray}
where the distribution amplitude $\phi_B$ is of the form
\begin{eqnarray}
\phi_B(x_B,b_B)= N_B x_B^2(1-x_B)^2
\mathrm{exp}\left[-\frac{(x_Bm_B)^2}{2\omega_{B}^2} -\frac{1}{2} (\omega_{B}b_B)^2\right],
\label{phib}
\end{eqnarray}
where $N_B$ is the normalization factor,  $\omega_B = 0.40 \pm 0.04$ GeV for $B^{\pm,0}$ and
$\omega_{B_s}=0.50 \pm 0.05$ for $B^0_s$, respectively.

For the $K\bar K$ system along with the subprocesses $\rho\to K\bar K$ and $\omega\to K\bar K$, the distribution
amplitudes are organized into~\cite{prd103.056021,prd91.094024}
\begin{eqnarray}
  \phi^{P\text{-wave}}_{K\bar K}(x,s)=\frac{-1}{\sqrt{2N_c}}
      \left[\sqrt{s}\,{\epsilon\hspace{-1.5truemm}/}\!_L\phi^0(x,s)
             + {\epsilon\hspace{-1.5truemm}/}\!_L {p\hspace{-1.7truemm}/} \phi^t(x,s)
             +\sqrt s \phi^s(x,s)  \right]\!,
\end{eqnarray}
with
\begin{eqnarray}
   \phi^{0}(x,s)&=&\frac{3C_X F_K(s)}{\sqrt{2N_c}} x(1-x)\left[1+a_R^{0} C^{3/2}_2(1-2x) \right]\!,\label{def-DA-0}\\
   \phi^{t}(x,s)&=&\frac{3C_X F^t_K(s)}{2\sqrt{2N_c}}(1-2x)^2\left[1+a_R^t  C^{3/2}_2(1-2x)\right]\!,\label{def-DA-t}\\
   \phi^{s}(x,s)&=&\frac{3C_X F^s_K(s)}{2\sqrt{2N_c}}(1-2x)\left[1+a_R^s\left(1-10x+10x^2\right) \right]\!,\label{def-DA-s}
\end{eqnarray}
where $F_K$ is the abbreviation of the vector time-like form factors in Eqs.~(\ref{def-F-u})-(\ref{def-F-ud}).
Moreover, we factor out the normalisation constant
\begin{eqnarray}
 C_{\rho^0}=C_{\omega}=\sqrt 2, \qquad C_{\rho^\pm}=1.
\end{eqnarray}
to make sure the the proper normalizations for the kaon time-like form factors.
The Gegenbauer moments have been catered to the data in Ref.~\cite{plb763-29} for the quasi-two-body decays
$B\to K\rho\to K\pi\pi$. Within flavour symmetry, we adopt the same Gegenbauer moments for the $P$-wave
$K\bar K$ system originating from the intermediate states $\omega$ and $\rho$ in this work. The vector time-like
form factors $F^t_K$ and $F^s_K$ for the twist-$3$ distribution amplitudes are deduced from the relations
$F^{t,s}_K(s)\approx (f^T_{\rho}/f_{\rho})F_K(s)$ and $F^{t,s}_K(s)\approx (f^T_{\omega}/f_{\omega})F_K(s)$~\cite{plb763-29}. The relation $f^T_\rho/f_\rho\approx f^T_\omega/f_\omega$~\cite{jhep1608-098} is
employed because of the lack of a lattice QCD determination for $f^T_\omega$, and the result $f^T_\rho/f_\rho=0.687$
is adopt at the scale $\mu=2$ GeV~\cite{prd78.114509}.

%%%%%%%%%%%%%%%%%%%%%%=FeynFig
\begin{figure}[tbp]
\centerline{\epsfxsize=14cm \epsffile{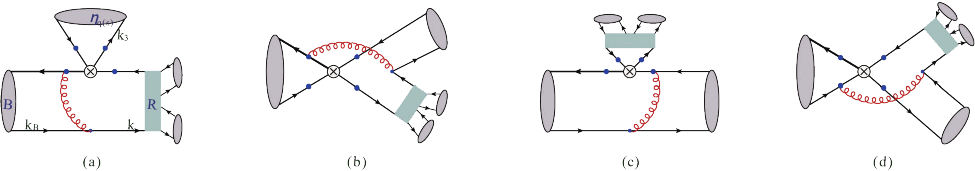}}
\caption{Typical Feynman diagrams for the processes $B\to\eta_{q(s)} R \to \eta_{q(s)} K\bar K $, with $R$ represents
               the resonances $\rho$, $\omega$ and their excited states. The dots on the quarks connecting the weak vertex
               $\otimes$ are the switchable vertices for the hard gluons.
              }
\label{fig-feyndiag}
\end{figure}
%%%%%%%%%%%%%%%%%%%%%%=FeynFig

With the subprocesses $\rho^+\to K^+ \bar K^0 $, $\rho^-\to K^- K^0$, $\rho^0 \to K^+ K^-$, $\rho^0 \to K^0\bar K^0$,
$\omega \to K^+ K^-$ and $\omega \to K^0\bar K^0$, and $\rho$ is $\rho(770), \rho(1450)$ or $\rho(1700)$ and $\omega$
is $\omega(782), \omega(1420)$ or $\omega(1650)$, the Lorentz invariant decay amplitudes for the quasi-two-body decays
$B\to \eta^{(\prime)}\rho\to \eta^{(\prime)}K\bar K $ and $B\to\eta^{(\prime)}\omega \to \eta^{(\prime)}K\bar K$ are given as follows:
\begin{eqnarray}
	{\cal A}(B^+ \to \eta_q\rho^+) &=& \frac{G_F} {2}
	\big\{V_{ub}^*V_{ud}[(C_1+\frac{C_2}{3})F^{LL}_{T\rho}+(\frac{C_1}{3}+C_2)(F^{LL}_{A\rho}+F^{LL}_{TP}+F^{LL}_{AP})\nonumber\\
	&+&C_2 M^{LL}_{T\rho}+C_1(M^{LL}_{A\rho}+M^{LL}_{TP}+M^{LL}_{AP})]-V_{tb}^*V_{td}[(C_5-\frac{C_7}{2})M^{LR}_{T\rho}\nonumber\\
	&+&(\frac{7 C_3}{3}+\frac{5 C_4}{3}+\frac{1}{3}(C_9-C_{10}))F^{LL}_{T\rho}+(2(C_5+\frac{C_6}{3})+\frac{1}{2}(C_7\nonumber\\
	&+&\frac{C_8}{3}))F^{LR}_{T\rho}+(\frac{C_5}{3}+C_6-\frac{1}{2}(\frac{C_7}{3}+C_8))F^{SP}_{T\rho}+(C_3+2C_4-\frac{C_9}{2}\nonumber\\
	&+&\frac{C_{10}}{2})M^{LL}_{T\rho}+(\frac{C_3}{3}+C_4+\frac{C_9}{3}+C_{10})(F^{LL}_{A\rho}+F^{LL}_{TP}+F^{LL}_{AP})
	+(\frac{C_5}{3}\nonumber\\
	&+&C_6+\frac{C_7}{3}+C_8)(F^{SP}_{A\rho}+F^{SP}_{AP}+F^{SP}_{TP})
	+(C_3+C_9)(M^{LL}_{A\rho}+M^{LL}_{TP}\nonumber\\
	&+& M^{LL}_{AP})
	+(C_5+C_7)(M^{LR}_{A\rho}+M^{LR}_{TP}+M^{LR}_{AP})+(2C_6+\frac{C_8}{2})M^{SP}_{T\rho}]\big\} \;,\label{amp15}
\end{eqnarray}
\begin{eqnarray}
	{\cal A}(B^+ \to \eta_s\rho^+) &=& \frac{G_F} {\sqrt{2}}
	\big\{-V_{tb}^*V_{td}[(C_3+\frac{C_4}{3}-\frac{1}{2}(C_9+\frac{C_{10}}{3}))F^{LL}_{T\rho}+(C_5+\frac{C_6}{3}\nonumber\\
	&+&\frac{1}{2}(C_7+\frac{C_8}{3}))F^{LR}_{T\rho}
	+(C_4-\frac{C_{10}}{2})M^{LL}_{T\rho}+(C_6-\frac{C_8}{2})M^{SP}_{T\rho}]\big\} \;,\label{amp16}
\end{eqnarray}
\begin{eqnarray}
	{\cal A}(B^+ \to \eta\rho^+) &=& {\cal A}(B^+ \to \eta_q\rho^+ ) \cos{\phi}-{\cal A}(B^+ \to\eta_s \rho^+ )\sin{\phi} \;,\label{amp17}\\
	{\cal A}(B^+ \to \eta^{\prime}\rho^+) &=&  {\cal A}(B^+ \to \eta_q\rho^+ )\sin{\phi}+ {\cal A}(B^+ \to  \eta_s\rho^+) \cos{\phi} \;,\label{amp18}
\end{eqnarray}
%------------------------------------------------------------------------------------------------
\begin{eqnarray}
	{\cal A}(B^0 \to \eta_q\rho^0) &=& -\frac{G_F} {2\sqrt{2}}
	\big\{V_{ub}^*V_{ud}[(C_1+\frac{C_2}{3})(F^{LL}_{T\rho}-F^{LL}_{A\rho}-F^{LL}_{TP}-F^{LL}_{AP})+C_2( M^{LL}_{T\rho}-M^{LL}_{A\rho}\nonumber\\
	&-&M^{LL}_{TP}-M^{LL}_{AP})]-V_{tb}^*V_{td}[(C_5-\frac{C_7}{2})M^{LR}_{T\rho}+(\frac{7 C_3}{3}+\frac{5 C_4}{3}+\frac{1}{3}(C_9-C_{10}))F^{LL}_{T\rho}\nonumber\\
	&-&(2(C_5+\frac{C_6}{3})+\frac{1}{2}(C_7+\frac{C_8}{3}))F^{LR}_{T\rho}+(\frac{C_5}{3}+C_6-\frac{1}{2}(\frac{C_7}{3}+C_8))(F^{SP}_{T\rho}-F^{SP}_{TP})\nonumber\\
	&+&(C_3+2C_4-\frac{C_9}{2}+\frac{C_{10}}{2})M^{LL}_{T\rho}-(-\frac{C_3}{3}-C_4+\frac{5 C_9}{3}+C_{10})(F^{LL}_{A\rho}+F^{LL}_{AP})\nonumber\\
	&+&\frac{3}{2}(C_7+\frac{C_8}{3})(F^{LR}_{A\rho}+F^{LR}_{AP})-(-\frac{C_5}{3}-C_6+\frac{1}{2}(\frac{C_7}{3}+C_8))(F^{SP}_{A\rho}+F^{SP}_{AP})\nonumber\\
	&+&(2C_6+\frac{C_8}{2})M^{SP}_{T\rho}-(-C_3+\frac{C_9}{2}+\frac{3 C_{10}}{2})(M^{LL}_{A\rho}+M^{LL}_{TP}+M^{LL}_{AP})\nonumber\\
	&-&(-C_5+\frac{C_7}{2})(M^{LR}_{A\rho}+M^{LR}_{TP}
	+M^{LR}_{AP})-\frac{3 C_8}{2}(M^{SP}_{A\rho}+M^{SP}_{TP}+M^{SP}_{AP})\nonumber\\
	&-&(-\frac{C_3}{3}-C_4+\frac{5 C_9}{3}+C_{10})F^{LL}_{TP}-\frac{3}{2}(C_7+\frac{C_8}{3})F^{LR}_{TP}]\big\}\;,\label{amp19}
\end{eqnarray}
%------------------------------------------------------------------------------------------------
\begin{eqnarray}
	{\cal A}(B^0 \to \eta_s\rho^0) &=& -\frac{G_F} {2}
	\big\{-V_{tb}^*V_{td}[(C_3+\frac{C_4}{3}-\frac{1}{2}(C_9+\frac{C_{10}}{3}))F^{LL}_{T\rho}+(-C_5-\frac{C_6}{3}\nonumber\\
	&+&\frac{1}{2}(C_7+\frac{C_8}{3}))F^{LR}_{T\rho}+(C_4-\frac{C_{10}}{2})M^{LL}_{T\rho}+(C_6-\frac{C_8}{2})M^{SP}_{T\rho}]\big\} \;,\label{amp20}
\end{eqnarray}
%------------------------------------------------------------------------------------------------
\begin{eqnarray}
	{\cal A}(B^0 \to \eta\rho^0) &=&{\cal A}(B^0 \to \eta_q\rho^0 ) \cos{\phi}-{\cal A}(B^0 \to \eta_s\rho^0 )\sin{\phi} \;,\label{amp21}\\
	{\cal A}(B^0 \to \eta^{\prime}\rho^0) &=& {\cal A}(B^0 \to \eta_q\rho^0 )\sin{\phi}+{\cal A}(B^0 \to \eta_s\rho^0 ) \cos{\phi} \;,\label{amp22}
\end{eqnarray}
%------------------------------------------------------------------------------------------------
\begin{eqnarray}
	{\cal A}(B_s^0 \to \eta_q\rho^0) &=& \frac{G_F} {2\sqrt{2}}
	\big\{V_{ub}^*V_{us}[(C_1+\frac{C_2}{3})(F^{LL}_{A\rho}+F^{LL}_{AP})+C_2(M^{LL}_{A\rho}+M^{LL}_{AP})]\nonumber\\
	&-&V_{tb}^*V_{ts}[\frac{3}{2}(C_9+\frac{C_{10}}{3})(F^{LL}_{A\rho}+F^{LL}_{AP})+\frac{3}{2}(C_7+\frac{C_8}{3})(F^{LR}_{A\rho}\nonumber\\
	&+&F^{LR}_{AP})+\frac{3 C_8}{2}(M^{SP}_{A\rho}+M^{SP}_{AP})+\frac{3 C_{10}}{2}(M^{LL}_{A\rho}+M^{LL}_{AP})]\big\} \;,
\end{eqnarray}
%------------------------------------------------------------------------------------------------
\begin{eqnarray}
	{\cal A}(B_s^0 \to \eta_s\rho^0) &=& \frac{G_F} {2}
	\big\{V_{ub}^*V_{us}[(C_1+\frac{C_2}{3})F^{LL}_{TP}+C_2M^{LL}_{TP}]-V_{tb}^*V_{ts}[\frac{3 C_8}{2}M^{SP}_{TP}\nonumber\\
	&+&\frac{3}{2}(C_7+\frac{C_8}{3})F^{LR}_{TP}+\frac{3}{2}(C_9+\frac{C_{10}}{3})F^{LL}_{TP}+\frac{3 C_{10}}{2}M^{LL}_{TP}]\big\} \;,
\end{eqnarray}
%------------------------------------------------------------------------------------------------
\begin{eqnarray}
	{\cal A}(B_s^0 \to \eta\rho^0) &=& {\cal A}(B_s^0 \to \eta_q\rho^0 ) \cos{\phi}-{\cal A}(B_s^0 \to \eta_s\rho^0 )\sin{\phi} \;,\\
	{\cal A}(B_s^0 \to \eta^{\prime}\rho^0) &=& {\cal A}(B_s^0 \to \eta_q\rho^0 )\sin{\phi}+{\cal A}(B_s^0 \to \eta_s\rho^0 ) \cos{\phi}  \;,
\end{eqnarray}
%------------------------------------------------------------------------------------------------
\begin{eqnarray}
	{\cal A}(B^0 \to \eta_q\omega) &=& \frac{G_F} {2\sqrt{2}}
	\big\{V_{ub}^*V_{ud}[(C_1+\frac{C_2}{3})(F^{LL}_{T\omega}+F^{LL}_{TP}+F^{LL}_{A\omega}+F^{LL}_{AP})+C_2(M^{LL}_{T\omega}+M^{LL}_{TP}\nonumber\\
	&+&M^{LL}_{A\omega}+M^{LL}_{AP})]-V_{tb}^*V_{td}[(\frac{7}{3}C_3+\frac{5}{3}C_4+\frac{C_9}{3}-\frac{C_{10}}{3})(F^{LL}_{T\omega}+F^{LL}_{TP})
	\nonumber\\
	&+&(2C_5+\frac{2 C_6}{3}+\frac{ C_7}{2}+\frac{ C_8}{6})(F^{LR}_{T\omega}+F^{LR}_{TP})
	+(\frac{ C_{5}}{3}+C_6-\frac{C_7}{6}-\frac{C_8}{2})(F^{SP}_{T\omega}\nonumber\\
	&+&F^{SP}_{TP})(C_3+2C_4-\frac{C_9}{2}+\frac{C_{10}}{2})(M^{LL}_{T\omega}+M^{LL}_{TP})+(C_5-\frac{C_7}{2})(M^{LR}_{T\omega}+M^{LR}_{TP})]\nonumber\\
	&+&(2C_6+\frac{C_8}{2})(M^{SP}_{T\omega}+M^{SP}_{TP})
	+(\frac{7C_3}{3}+\frac{5C_4}{3}-\frac{C_9}{3}-\frac{C_{10}}{3})(F^{LL}_{A\omega}+F^{LL}_{AP})\nonumber\\
	&+&(2C_5+\frac{2C_6}{3}+\frac{C_7}{2}+\frac{C_8}{6})(F^{LR}_{A\omega}+F^{LR}_{AP})+(\frac{C_5}{3}+C_6-\frac{C_7}{6}-\frac{C_8}{2})(F^{SP}_{A\omega}\nonumber\\
	&+&F^{SP}_{AP})+(C_3+2C_4-\frac{C_9}{2}+\frac{C_{10}}{2})(M^{LL}_{A\omega}+M^{LL}_{AP})+(C_5-\frac{C_7}{2})(M^{LR}_{A\omega}\nonumber\\
	&+&M^{LR}_{AP})+(2C_6+\frac{C_8}{2})(M^{SP}_{A\omega}+M^{SP}_{AP})]
	\big\} \;
\end{eqnarray}
%------------------------------------------------------------------------------------------------
\begin{eqnarray}
	{\cal A}(B^0 \to \eta_s\omega) &=& -\frac{G_F} {2}
	V_{tb}^*V_{td}[(C_3+\frac{C_4}{3}-\frac{C_9}{2}-\frac{C_{10}}{6})F^{LL}_{T\omega}
	+(C_5+\frac{C_6}{3}-\frac{C_7}{2}-\frac{C_8}{6})F^{LR}_{T\omega}\nonumber\\
	&+&(C_4-\frac{C_{10}}{2})M^{LL}_{T\omega}+(C_6-\frac{C_8}{2})M^{SP}_{T\omega}] \;,
\end{eqnarray}
%------------------------------------------------------------------------------------------------
\begin{eqnarray}
	{\cal A}(B^0 \to \eta\omega) &=& {\cal A}(B^0 \to \eta_q\omega ) \cos{\phi}-{\cal A}(B^0 \to \eta_s\omega )\sin{\phi} \;,\\
	{\cal A}(B^0 \to \eta^{\prime}\omega) &=& {\cal A}(B^0 \to \eta_q\omega )\sin{\phi}+{\cal A}(B^0 \to \eta_s\omega ) \cos{\phi}  \;,
\end{eqnarray}
%------------------------------------------------------------------------------------------------
\begin{eqnarray}
	{\cal A}(B_s^0 \to \eta_q\omega) &=& \frac{G_F} {2\sqrt{2}}
	\big\{V_{ub}^*V_{us}[(C_1+\frac{C_2}{3})(F^{LL}_{A\omega}+F^{LL}_{AP})+C_2(M^{LL}_{A\omega}+M^{LL}_{AP})]
	-V_{tb}^*V_{ts}\nonumber\\&\times&[
	(2C_4+\frac{ C_{10}}{2})(M^{LL}_{A\omega}+M^{LL}_{AP})+(2C_3+\frac{2C_4}{3}+\frac{C_9}{2}+\frac{C_{10}}{6})(F^{LL}_{A\omega}+F^{LL}_{AP})\nonumber\\
	&+&(2C_5+\frac{ 2C_{6}}{3}+\frac{C_7}{2}+\frac{C_8}{6})(F^{LR}_{A\omega}+F^{LR}_{AP})\nonumber\\
	&+&(2C_6+\frac{C_8}{2})(M^{SP}_{A\omega}+M^{SP}_{AP})]
	\big\} \;,
\end{eqnarray}
%------------------------------------------------------------------------------------------------
\begin{eqnarray}
	{\cal A}(B_s^0 \to \eta_s\omega) &=& \frac{G_F} {2}
	\big\{V_{ub}^*V_{us}[(C_1+\frac{C_2}{3})F^{LL}_{TP}+C_2M^{LL}_{TP}]
	-V_{tb}^*V_{ts}[(2C_3+\frac{2C_4}{3}+\frac{C_9}{2}
	\nonumber\\
	&+&\frac{C_{10}}{6})F^{LL}_{TP}+(2C_5+\frac{2C_6}{3}+\frac{C_7}{2}+\frac{C_8}{6})F^{LR}_{TP}
	+(2C_4+\frac{ C_{10}}{2})M^{LL}_{TP}\nonumber\\
	&+&(2C_6+\frac{ C_{8}}{2})M^{SP}_{TP}]
	\big\} \;
\end{eqnarray}
%------------------------------------------------------------------------------------------------
\begin{eqnarray}
	{\cal A}(B_s^0 \to \eta\omega) &=& {\cal A}(B_s^0 \to \eta_q\omega ) \cos{\phi}-{\cal A}(B_s^0 \to \eta_s\omega )\sin{\phi} \;,\\
	{\cal A}(B_s^0 \to \eta^{\prime}\omega) &=& {\cal A}(B_s^0 \to \eta_q\omega )\sin{\phi}+{\cal A}(B_s^0 \to \eta_s\omega ) \cos{\phi}  \;,
	\label{amp26}
\end{eqnarray}
where $G_F$ is the Fermi coupling constant.  $V_{ij}$'s are the CKM matrix elements. %abibbo-Kobayashi-Maskawa

The general amplitudes for the concerned quasi-two-body decays in the Eqs.~(\ref{amp15})-(\ref{amp26}) are deduced
according to Fig.~\ref{fig-feyndiag}. For the emission diagrams the Fig.~\ref{fig-feyndiag}-(a), one has
\begin{eqnarray}
	F^{LL}_{T\rho(\omega)} &=& 8\pi C_F m^4_B f_h (\zeta-1)\int dx_B dz\int b_B db_B b db \phi_B(x_B,b_B)\big\{\big[(1+z)\phi^0+\sqrt{\zeta}(1
	-2z)\nonumber\\
	&\times&(\phi^s+\phi^t)\big]
    E_{a12}(t_{a1})h_{a1}(x_B,z,b_B,b)+[\zeta \phi^0+2 \sqrt{\zeta} \phi^s]E_{a12}(t_{a2})\nonumber\\
    &\times&h_{a2}(x_B,z,b_B,b) \big\}\;,
	\label{def-Fll-Tphi}\\
	%----------------
	F^{LR}_{T\rho(\omega)} &=& -F^{LL}_{T\rho(\omega)}\;, \\
%\end{eqnarray}
%----------------
%\begin{eqnarray}
	F^{SP}_{T\rho(\omega)} &=& 16\pi C_F m^4_B r f_h \int dx_B dz\int b_B db_B b db \phi_B(x_B,b_B)\big\{\big[(\zeta(2z-1)+1)\phi^0+\sqrt{\zeta}((2\nonumber\\&+&z)\phi^s-z\phi^t)\big]E_{a12}(t_{a1})h_{a1}(x_B,z,b_B,b)+\big[x_B\phi^0+2\sqrt{\zeta}(\zeta-x_B+1)\phi^s\big]
	\nonumber\\
	&\times& E_{a12}(t_{a2})h_{a2}(x_B,z,b_B,b)\big\}\;,
\end{eqnarray}
	%----------------
\begin{eqnarray}
		M^{LL}_{T\rho(\omega)} &=& 32\pi C_F m^4_B/\sqrt{2N_c} (\zeta-1)\int dx_B dz dx_3\int b_B db_B b_3 db_3\phi_B(x_B,b_B)\phi^A\big\{\big[((1-\zeta)\nonumber\\
		&\times&(1-x_3)-x_B-z\zeta)\phi^0-\sqrt{\zeta}z(\phi^s-\phi^t)\big] E_{a34}(t_{a3})h_{a3}(x_B,z,x_3,b_B,b_3)+\big[(x_3(\zeta\nonumber\\&-&1)+x_B-z)\phi^0+z\sqrt{\zeta}(\phi^s +\phi^t)\big]E_{a34}(t_{a4})h_{a4}(x_B,z,x_3,b_B,b_3) \big\}\;,
\end{eqnarray}

\begin{eqnarray}
		M^{LR}_{T\rho(\omega)} &=& 32\pi C_F r m^4_B/\sqrt{2N_c}\int dx_B dz dx_3\int b_B db_B b_3 db_3\phi_B(x_B,b_B)\big\{\big[((1-x_3)(1-\zeta)\nonumber\\
		&-&x_B)(\phi^P+\phi^T)(\phi^0+\sqrt{\zeta}(\phi^s-\phi^t))-\sqrt{\zeta} z(\phi^P-\phi^T)(\sqrt{\zeta}\phi^0 -\phi^s-\phi^t)\nonumber\\
		&\times&E_{a34}(t_{a3})h_{a3}(x_B,z,x_3,b_B,b_3)+\big[\sqrt{\zeta}z(\phi^P+\phi^T)(\sqrt{\zeta}\phi^0-\phi^s-\phi^t) +(x_B-x_3(1\nonumber\\
		&-&\zeta))(\phi^P-\phi^T)(\phi^0+\sqrt{\zeta}(\phi^s-\phi^t))\big]E_{a34}(t_{a4})h_{a4}(x_B,z,x_3,b_B,b_3)\big\}\;,\\
		%----------------
		M^{SP}_{T\rho(\omega)} &=& 32\pi C_F m^4_B/\sqrt{2N_c} (\zeta-1)\int dx_B dz dx_3\int b_B db_B b_3 db_3\phi_B(x_B,b_B)\phi^A\big\{\big[((1-\zeta)(x_3\nonumber\\
		&-&1)+x_B-z)\phi^0+\sqrt{\zeta}z(\phi^s+\phi^t)\big]E_{a34}(t_{a3})h_{a3}(x_B,z,x_3,b_B,b_3)+\big[(x_3(1-\zeta)-x_B\nonumber\\
		&-&z\zeta)\phi^0-z\sqrt{\zeta}(\phi^s-\phi^t) \big]E_{a34}(t_{a4})h_{a4}(x_B,z,x_3,b_B,b_3) \big\}\;,
\end{eqnarray}
where  the color factor $C_F = 4/3$ and the ratio $r = m_0^h/m_B$. The symbols $LL$, $LR$ and $SP$ are employed to
denote the amplitudes from the $(V-A)(V-A)$, $(V-A)(V+A)$ and $(S-P)(S+P)$ operators, respectively.
For the factorizable diagrams in Fig.~\ref{fig-feyndiag}, we name their expressions with $F$, while the others are nonfactorizable diagrams, we name their expressions with $M$.
The annihilation-type diagrams Fig.~\ref{fig-feyndiag}-(b) give us
\begin{eqnarray}
		F^{LL}_{A\rho(\omega)} &=& 8\pi C_F m^4_B f_B\int dz dx_3\int b db b_3 db_3 \big\{\big[(1-\zeta)(1-z)\phi^A\phi^0+2r\sqrt{\zeta}\phi^P((z-2)\phi^s\nonumber\\
		&-&z\phi^t)\big]E_{b12}(t_{b1})h_{b1}(z,x_3,b,b_3)+\big[[(1-x_3)\zeta^2+(2x_3-1)\zeta-x_3]\phi^A\phi^0+2r\sqrt{\zeta}[((1\nonumber\\
		&-&x_3)\zeta+x_3)(\phi^P+\phi^T)+(\phi^P-\phi^T)]\phi^s\big]E_{b12}(t_{b2})h_{b2}(z,x_3,b,b_3) \big\}\;,\\
		%----------------
		F^{LR}_{A\rho(\omega)} &=& -F^{LL}_{A\rho(\omega)}\;,\\
		%----------------
		F^{SP}_{A\rho(\omega)} &=&  16\pi C_F m^4_B f_B \int dz dx_3\int b db b_3 db_3 \big\{\big[2r(1+(z-1)\zeta)\phi^P\phi^0-\sqrt{\zeta}(1-\zeta)(1-z)\nonumber\\
		&\times&\phi^A(\phi^s+\phi^t)\big] E_{b12}(t_{b1})h_{b1}(z,x_3,b,b_3)+[r\left(x_3(1-\zeta)(\phi^P-\phi^T)-2\zeta \phi^T\right)\phi^0+2\sqrt{\zeta}\nonumber\\
		&\times&(\zeta-1)\phi^A\phi^s ]E_{b12}(t_{b2}) h_{b2}(z,x_3,b,b_3)\big\}\;,
\end{eqnarray}

\begin{eqnarray}	
		M^{LL}_{A\rho(\omega)} &=& 32\pi C_F m^4_B/\sqrt{2N_c} \int dx_B dz dx_3\int b_B db_B b db\phi_B(x_B,b_B)\big\{\big[[(x_3-z-1)\zeta^2+(1
		\nonumber\\
		&+&z-2 x_3-x_B)\zeta+x_3+x_B]\phi^A\phi^0+ r\sqrt\zeta[z(\phi^P-\phi^T)(\phi^s+\phi^t)+((1-x_3)(1-\zeta)\nonumber\\
		&-&x_B)(\phi^P+\phi^T)(\phi^s-\phi^t)-4\phi^P\phi^s]\big] E_{b34}(t_{b3})h_{b3}(x_B,z,x_3,b_B,b)+\big[(1-\zeta)^2(z-1)\nonumber\\
		&\times&\phi^A\phi^0+ r\sqrt\zeta [(\zeta(1-x_3)+x_3-x_B)(\phi^P-\phi^T)(\phi^s+\phi^t)+ (1-z)(\phi^P+\phi^T)\nonumber\\
		&-&(\phi^s-\phi^t)]\big]E_{b34}(t_{b4})h_{b4}(x_B,z,x_3,b_B,b)\big\}\;,\\
		M^{LR}_{A\rho(\omega)} &=& 32\pi C_F m^4_B/\sqrt{2N_c} \int dx_B dz dx_3\int b_B db_B b db \phi_B(x_B,b_B)\big\{\big[r [(2+\zeta x_3-x_3-x_B)\nonumber\\
		&\times&(\phi^P+\phi^T)-\zeta z(\phi^P-\phi^T)-2 \zeta \phi^P]\phi^0+\sqrt\zeta(1-\zeta)(1+z)\phi^A(\phi^s-\phi^t)\big] E_{b34}(t_{b3})\nonumber\\
		&\times& h_{b3}(x_B,z,x_3,b_B,b)+ \big[r[(x_3(1-\zeta)-x_B)(\phi^P+\phi^T)+\zeta z(\phi^P-\phi^T)+2\zeta \phi^T]\phi^0+\nonumber\\
		&+&\sqrt\zeta(1-\zeta)(1-z)\phi^A(\phi^s-\phi^t) \big]E_{b34}(t_{b4})h_{b4}(x_B,z,x_3,b_B,b)\big\}\;,\\
		M^{SP}_{A\rho(\omega)} &=& 32\pi C_F m^4_B/\sqrt{2N_c} \int dx_B dz dx_3\int b_B db_B b db\phi_B(x_B,b_B)\big\{\big[(\zeta-1)[(\zeta-1)z+1]\phi^A\phi^0\nonumber\\
		&+& r\sqrt\zeta [((1-\zeta)(x_3-1)+x_B)(\phi^P-\phi^T)(\phi^s+\phi^t)-z(\phi^P+\phi^T)(\phi^s-\phi^t)+4\phi^P\phi^s]\big]\nonumber\\
		&\times&E_{b34}(t_{b3})h_{b3}(x_B,z,x_3,b_B,b)+ \big[[(\zeta-1)(x_3(\zeta-1)+x_B)+\zeta z(1-\zeta)]\phi^A\phi^0+r\sqrt\zeta\nonumber\\
		&\times& [(z-1)(\phi^P-\phi^T)(\phi^s+\phi^t)+((\zeta-1)x_3+x_B-\zeta)(\phi^P+\phi^T)(\phi^s-\phi^t)]\big] \nonumber\\
		&\times&E_{b34}(t_{b4})h_{b4}(x_B,z,x_3,b_B,b)\big\}\;,
\end{eqnarray}
		
The amplitudes from the emission diagrams the Fig.~\ref{fig-feyndiag}-(c) are written as	
\begin{eqnarray}		
		F^{LL}_{TP} &=& 8\pi C_F m^4_B F_K\int dx_B dx_3\int b_B db_B b_3 db_3 \phi_B(x_B,b_B)\big\{\big[(1-\zeta)[(x_3(\zeta-1)-1)\phi^A
		+ r \nonumber\\
		&\times&(2x_3-1)
	   \phi^P]-r(1+\zeta-2x_3(1-\zeta))\phi^T\big]E_{c12}(t_{c1})h_{c1}(x_B,x_3,b_B,b_3)+[x_B(1-\zeta) \nonumber\\
	   &\times&\zeta\phi^A-2r(1-\zeta(1-x_B))\phi^P] E_{c12}(t_{c2})
		 h_{c2}(x_B,x_3,b_B,b_3)\big\}\;,\\
		F^{LR}_{TP} &=& F^{LL}_{TP},
\end{eqnarray}

\begin{eqnarray}
		M^{LL}_{TP} &=& 32\pi C_F m^4_B/\sqrt{2N_c} \int dx_B dz dx_3\int b_B db_B b db
		\phi_B(x_B,b_B)\phi^0 \big\{\big[(x_B+z-1)(1-\zeta)^2\nonumber\\
		&\times&\phi^A +r[\zeta(x_B+z)
		\times(\phi^P+\phi^T) +x_3(1-\zeta)(\phi^P-\phi^T)-2\zeta\phi^T]\big] E_{c34}(t_{c3})\nonumber\\
		&\times&h_{c3}(x_B,z,x_3,b_B,b)+\big[(\zeta-1)[x_3(\zeta-1)+x_B
		-z]\phi^A+r[x_3(\zeta-1)(\phi^P+\phi^T)\nonumber\\&-&(x_B-z)\zeta(\phi^P_K-\phi^T_K)]\big]\times E_{c34}(t_{c4})h_{c4}(x_B,z,x_3,b_B,b)\big\}\;,\\
		M^{LR}_{TP} &=& 32\pi C_F m^4_B\sqrt{\zeta}/\sqrt{2N_c} \int dx_B dz dx_3\int b_B db_B b db \phi_B(x_B,b_B)\big\{\big[(1-x_B-z)(\zeta-1)\nonumber\\
		&\times&(\phi^s+\phi^t)\phi^A-r
		\times(x_3(1-\zeta)+\zeta)(\phi^s-\phi^t)(\phi^P+\phi^T)-r(1-x_B-z)(\phi^s+\phi^t)\nonumber\\
		&\times&(\phi^P-\phi^T)\big]E_{c34}(t_{c3})h_{c3}(x_B,z,x_3,b_B,b)
		+\big[(z-x_B)(1-\zeta)(\phi^s-\phi^t)\phi^A+rx_3(1\nonumber\\&-&\zeta)(\phi^s+\phi^t)(\phi^P+\phi^T)+r(z-x_B)(\phi^s-\phi^t)(\phi^P-\phi^T)\big]
		\nonumber\\
		&\times&E_{c34}(t_{c4})h_{c4}(x_B,z,x_3,b_B,b)\big\}\;,\\
		M^{SP}_{TP} &=& 32\pi C_F m^4_B/\sqrt{2N_c} \int dx_B dz dx_3\int b_B db_B b db \phi_B(x_B,b_B)\phi^0\big\{\big[(\zeta(x_3-1)-x_3+x_B\nonumber\\&+&z-1)
		(1-\zeta)\phi^A
		+r x_3(1-\zeta)(\phi^P+\phi^T)+ r\zeta(x_B+z)(\phi^P-\phi^T)+ 2r\zeta\phi^T\big] E_{c34}(t_{c3})\nonumber\\
		&\times&
		h_{c3}(x_B,z,x_3,b_B,b)+\big[(z-x_B)
		\times(1-\zeta)^2\phi^A+r\zeta(z-x_B)(\phi^P+\phi^T)
		\nonumber\\&-&rx_3(1-\zeta)(\phi^P-\phi^T)\big]E_{c34}(t_{c4})h_{c4}(x_B,z,x_3,b_B,b)\big\}\;,
			\end{eqnarray}
			
The annihilation-type diagrams Fig.~\ref{fig-feyndiag}-(d) contribute the following amplitudes:		
\begin{eqnarray}
 F^{LL}_{AP} &=& 8\pi C_F m^4_B f_B \int dz dx_3\int b db b_3 db_3\big\{\big [(x_3(1-\zeta)-1)(\zeta-1)\phi^0\phi^A+ 2r\sqrt\zeta\phi^s[x_3(\zeta-1)
	\nonumber\\
	&\times&(\phi^P-\phi^T) +2\phi^P]\big]
		\times E_{d12}(t_{d1})h_{d1}(z,x_3,b,b_3)+\big[z(\zeta-1)\phi^0\phi^A-2r\sqrt\zeta[z(\phi^s+\phi^t)
			\nonumber\\&+&(1-\zeta)(\phi^s-\phi^t)]\phi^P\big] E_{d12}(t_{d2})
		\times h_{d2}(z,x_3,b,b_3) \big\}\;,\\
		F^{LR}_{AP} &=& -F^{LL}_{AP}\;,\\		
		F^{SP}_{AP} &=& 16\pi C_F m^4_B f_B  \int dz dx_3\int b db b_3 db_3 \big\{\big[(\zeta-1)[2\sqrt\zeta\phi^s\phi^A+r(1-x_3)\phi^0 \phi^P]- r[\zeta+x_3\nonumber\\
		&\times&(\zeta-1)+1]\phi^0\phi^T\big]
		\times E_{d12}(t_{d1})h_{d1}(z,x_3,b,b_3)+ \big[z\sqrt\zeta(\zeta-1)(\phi^s-\phi^t)\phi^A+2r(z \zeta\nonumber\\&+&\zeta-1)\phi^0\phi^P \big] E_{d12}(t_{d2})h_{d2}(z,x_3,b,b_3)\big\}\;,
\end{eqnarray}

 \begin{eqnarray}
			M^{LL}_{AP} &=& 32\pi C_F m^4_B/\sqrt{2N_c} \int dx_B dz dx_3\int b_B db_B b db\phi_B(x_B,b_B) \big\{\big[[(x_B+z-1)\zeta^2+ (1-2x_B
			\nonumber\\&-& 2z)\zeta+x_B
			+z]\phi^0\phi^A- r\sqrt\zeta[(\eta(1-x_3)+x_3)(\phi^s-\phi^t)(\phi^P+\phi^T)+ (1-x_B-z)(\phi^s\nonumber\\&+&\phi^t)
	    	(\phi^P-\phi^T)- 4\phi^s\phi^P]\big]
			E_{d34}(t_{d3})h_{d3}(x_B,z,x_3,b_B,b)+ \big[(\zeta-1)((1-x_3)(1-\zeta)+\zeta\nonumber\\&\times&(x_B-z))
			\phi^0\phi^A+r\sqrt\zeta[(x_B-z)(\phi^s-\phi^t) (\phi^P+\phi^T)+(1-\zeta)(x_3-1)(\phi^s+\phi^t)(\phi^P\nonumber\\&-&\phi^T)] \big]E_{d34}(t_{d4})h_{d4}(x_B,z,x_3,b_B,b) \big\}\;,\\
			M^{LR}_{AP} &=&  32\pi C_F m^4_B/\sqrt{2N_c} \int dx_B dz dx_3\int b_B db_B b db \phi_B(x_B,b_B)\big\{\big[\sqrt\zeta(1-\zeta)(x_B+z-2)(\phi^s\nonumber\\
			&+&\phi^t)\phi^A+ r\phi^0[\zeta (x_B+z-1)(\phi^P+\phi^T)+ (1+x_3-\zeta x_3)(\phi^P-\phi^T)-2\zeta \phi^T]\big]E_{d34}(t_{d3})\nonumber\\&\times&h_{d3}(x_B,z,x_3,b_B,b)+\big[\sqrt\zeta(1-\zeta)(x_B-z)(\phi^s+\phi^t)\phi^A+r\phi^0[\zeta(x_B-z)(\phi^P+\phi^T)\nonumber\\
			&+&(1-\zeta)(1-x_3)(\phi^P-\phi^T)]\big]E_{d34}(t_{d4})h_{d4}(x_B,z,x_3,b_B,b)\big\}\;,\\
			M^{SP}_{AP} &=& 32\pi C_F m^4_B/\sqrt{2N_c} \int dx_B dz dx_3\int b_B db_B b db\phi_B(x_B,b_B)\big\{\big[(\zeta-1)[x_3(\zeta-1)-\zeta(x_B\nonumber\\
			&+&z)+1]\phi^0\phi^A
			- r\sqrt\zeta[(x_B+z-1)(\phi^s-\phi^t)(\phi^P+\phi^T)+ (\zeta x_3-\zeta-x_3)(\phi^s+\phi^t)(\phi^P\nonumber\\
			&-&\phi^T)
			+14\phi^s\phi^P]\big]E_{d34}(t_{d3})
			\times h_{d3}(x_B,z,x_3,b_B,b)+ \big[(\zeta-1)^2(z-x_B)\phi^0\phi^A- r\sqrt\zeta[(1\nonumber\\
			&-&\zeta)(x_3-1)
			(\phi^s-\phi^t)(\phi^P+\phi^T)+ (x_B-z)(\phi^s+\phi^t)(\phi^P-\phi^T) \big] \nonumber\\
			&\times&E_{d34}(t_{d4})h_{d4}(x_B,z,x_3,b_B,b) \big\}\;. \label{def-Msp-ah}
		\end{eqnarray}
		
The PQCD functions $t_{xi}, h_{xi}$ and $E_{x12,x34}$, with $x\in\{a,b,c,d\}$ and $i\in\{1,2,3,4\}$,  appear in the factorization formulas, Eqs. (\ref{def-Fll-Tphi})-(\ref{def-Msp-ah}), have their explicit expressions in Appendix B of~\cite{jhep2003-162}.		

%%~~~~~~~~~~~~~~~ Refs ~~~~~~~~~~~~~~~%%

\end{document}